\documentclass[twocolumn]{aastex63}
\usepackage{multirow}
\usepackage{lineno}
\usepackage{float}
\usepackage{graphicx}
\usepackage{lipsum}
\usepackage{amsmath}

\begin{document}
\title{Hints of planet formation signatures in a large-cavity disk \\
studied in the AGE-PRO ALMA Large Program}

\correspondingauthor{Anibal Sierra}
\email{anibalsierram@gmail.com}

\author[0000-0002-5991-8073]{Anibal Sierra}
\affiliation{Departamento de Astronom\'ia, Universidad de Chile, Camino El Observatorio 1515, Las Condes, Santiago, Chile}
\affiliation{Mullard Space Science Laboratory, University College London, Holmbury St Mary, Dorking, Surrey RH5 6NT, UK}

\author[0000-0002-1199-9564]{Laura M. P\'erez}
\affiliation{Departamento de Astronom\'ia, Universidad de Chile, Camino El Observatorio 1515, Las Condes, Santiago, Chile}

\author[0000-0002-7238-2306]{Carolina Agurto-Gangas}
\affiliation{Departamento de Astronom\'ia, Universidad de Chile, Camino El Observatorio 1515, Las Condes, Santiago, Chile}

\author[0000-0002-1575-680X]{James Miley}
\affiliation{Departamento de Física, Universidad de Santiago de Chile, Av. Victor Jara 3659, Santiago, Chile}
\affiliation{Millennium Nucleus on Young Exoplanets and their Moons (YEMS), Chile}
\affiliation{Center for Interdisciplinary Research in Astrophysics Space Exploration (CIRAS), Universidad de Santiago, Chile}

\author[0000-0002-0661-7517]{Ke Zhang}
\affiliation{Department of Astronomy, University of Wisconsin-Madison, 475 N Charter St, Madison, WI 53706}

\author[0000-0001-8764-1780]{Paola Pinilla}
\affiliation{Mullard Space Science Laboratory, University College London, Holmbury St Mary, Dorking, Surrey RH5 6NT, UK}

\author[0000-0001-7962-1683]{Ilaria Pascucci}
\affiliation{Lunar and Planetary Laboratory, The University of Arizona, Tucson, AZ 85721, USA}

\author[0000-0002-8623-9703]{Leon Trapman}
\affiliation{Department of Astronomy, University of Wisconsin-Madison, 475 N Charter St, Madison, WI 53706}

\author[0000-0002-2358-4796]{Nicolas Kurtovic}
\affiliation{Max-Planck-Institut fur Astronomie (MPIA), Konigstuhl 17, 69117 Heidelberg, Germany}

\author[0000-0002-4147-3846]{Miguel Vioque}
\affiliation{European Southern Observatory, Karl-Schwarzschild-Str. 2, 85748 Garching bei München, Germany}
\affiliation{Joint ALMA Observatory, Avenida Alonso de Córdova 3107, Vitacura, Santiago, Chile}

\author[0000-0003-0777-7392]{Dingshan Deng}
\affiliation{Lunar and Planetary Laboratory, The University of Arizona, Tucson, AZ 85721, USA}

\author[0009-0004-8091-5055]{Rossella Anania}
\affiliation{Dipartimento di Fisica, Università degli Studi di Milano, Via Celoria 16, I-20133 Milano, Italy}

\author[0000-0003-2251-0602]{John Carpenter}
\affiliation{Joint ALMA Observatory, Avenida Alonso de Córdova 3107, Vitacura, Santiago, Chile}

\author[0000-0002-2828-1153]{Lucas A. Cieza}
\affiliation{Instituto de Estudios Astrofísicos, Universidad Diego Portales, Av. Ejercito 441, Santiago, Chile}

\author[0000-0003-4907-189X]{Camilo Gonz\'alez-Ruilova}
\affiliation{Instituto de Estudios Astrofísicos, Universidad Diego Portales, Av. Ejercito 441, Santiago, Chile}
\affiliation{Millennium Nucleus on Young Exoplanets and their Moons (YEMS), Chile}
\affiliation{Center for Interdisciplinary Research in Astrophysics Space Exploration (CIRAS), Universidad de Santiago, Chile}

\author[0000-0001-5217-537X]{Michiel Hogerheijde}
\affiliation{Leiden Observatory, Leiden University, PO Box 9513, 2300 RA Leiden, the Netherlands}
\affiliation{Anton Pannekoek Institute for Astronomy, University of Amsterdam, the Netherlands}

\author[0000-0002-6946-6787]{Aleksandra Kuznetsova}
\affiliation{Center for Computational Astrophysics, Flatiron Institute, 162 Fifth Ave., New York, New York, 10025}

\author[0000-0003-4853-5736]{Giovanni P. Rosotti}
\affiliation{Dipartimento di Fisica, Università degli Studi di Milano, Via Celoria 16, I-20133 Milano, Italy}

\author[0000-0003-3573-8163]{Dary A. Ruiz-Rodriguez}
\affiliation{National Radio Astronomy Observatory; 520 Edgemont Rd., Charlottesville, VA 22903, USA}
\affiliation{Joint ALMA Observatory, Avenida Alonso de Córdova 3107, Vitacura, Santiago, Chile}

\author[0000-0002-6429-9457]{Kamber Schwarz}
\affiliation{Max-Planck-Institut fur Astronomie (MPIA), Konigstuhl 17, 69117 Heidelberg, Germany}

\author[0000-0002-1103-3225]{Benoît Tabone}
\affiliation{Université Paris-Saclay, CNRS, Institut d'Astrophysique Spatiale, Orsay, France}

\author[0000-0001-9961-8203]{Estephani E. TorresVillanueva}
\affiliation{Department of Astronomy, University of Wisconsin-Madison, 475 N Charter St, Madison, WI 53706}

\begin{abstract}
Detecting planet signatures in protoplanetary disks is fundamental to understanding how and where planets form. In this work, we report dust and gas observational hints of planet formation in the disk around 2MASS-J16120668-301027, as part of the ALMA Large Program ``AGE-PRO: ALMA survey of Gas Evolution in Protoplanetary disks''. The disk was imaged with the Atacama Large Millimeter/submillimeter Array (ALMA) at Band 6 (1.3 mm) in dust continuum emission and four molecular lines: $^{12}$CO(J=2-1), $^{13}$CO(J=2-1),  C$^{18}$O(J=2-1), and H$_2$CO(J=3$_{(3,0)}$-2$_{(2,0)}$). 
Resolved observations of the dust continuum emission (angular resolution of $\sim 150$ mas, 20 au) show a ring-like structure with a peak at $0.57 ^{\prime \prime}$ (75 au), a deep gap with a minimum at 0.24$^{\prime \prime}$ (31 au), an inner disk, a bridge connecting the inner disk and the outer ring, along with a spiral arm structure, and a tentative detection (to $3\sigma$) of a compact emission at the center of the disk gap, with an estimated dust mass of $\sim 2.7-12.9$ Lunar masses.  We also detected a kinematic kink (not coincident with any dust substructure) through several $^{12}$CO channel maps (angular resolution $\sim$ 200 mas, 30 au), located at a radius of $\sim 0.875^{\prime \prime}$ (115.6 au).
After modeling the $^{12}$CO velocity rotation around the protostar, we identified a tentative rotating-like structure at the kink location with a geometry similar to that of the disk.
We discuss potential explanations for the dust and gas substructures observed in the disk, and their potential connection to signatures of planet formation.
\end{abstract}


\section{Introduction} \label{sec:introduction}

Planet formation takes place in protoplanetary disks \citep{Williams_2011}. Dust grains in disks (inherited from the original molecular cloud with sizes $\lesssim 1 \ \mu$m) collide, aggregate, and eventually form the cores of planets like Earth (thousands of kilometers in size) \citep{Birnstiel_2023}. The process of how dust grains can grow more than 12 orders of magnitude in size in less than a few million years remains a topic of debate. 

The different stages of planetary growth are not fully understood \citep{Drkazkowska_2023}, making observational evidence crucial to discern when and where planet formation takes place in protoplanetary disks. The meter-size barrier and the expected fast radial drift of solids \citep{Goldreich_1973, Weidenschilling_1977} are, to date, still one of the most challenging problems to understand in planet formation theory.

Observed dust continuum morphologies such as rings, gaps, vortices, or spirals at millimeter wavelengths \citep[e.g.,][]{Andrews_2020} have been interpreted as an indirect evidence of early planet formation around Class II disks\footnote{This classification pertains to young stellar objects categorized based on the shape of their Spectral Energy Distribution (SED) at infrared wavelengths. Specifically, Class II disks are characterized by a spectral index between 2 and 20 microns ranging from $-1.5 \lesssim \alpha_{IR} \lesssim 0$ \citep{Lada_1987}.}, although other processes can also explain these features \citep[e.g.,][]{Bae_2023}. 
Annular structures have been also found at millimeter observations in the young disks around IRS 63  \citep[a embedded Class I disk,][]{Segura-Cox_2020}, HL Tau \citep[a disk evolving from SED Class I to Class II,][]{ALMA_2015}, and in some of the Class I disks in the eDisks ALMA Large Program \citep{Ohashi_2023}. These findings suggest that planets could form, even at tens of au from their host stars, earlier than previously expected (within $\lesssim 1$ Myr).

Although the origin of the observed dust sub-structures has been associated with dust traps \citep[e.g.,][]{Pinilla_2012b, Birnstiel_2012} in several disks, it remains a subject of ongoing debate. Various physical mechanisms, including planet-disk interactions \citep[e.g.,][]{Dipierro_2017}, dead zones \citep[e.g.,][]{Regaly_2012, Flock_2015, Garate_2021}, traffic jams \citep[e.g.,][]{Okuzumi_2016}, snowlines \citep[e.g.,][]{Zhang_2015}, and photo-evaporation \citep{Alexander_2014, Garate_2023}, have been also proposed to explain the ring structures observed in protoplanetary disks \citep[e.g.,][]{Andrews_2018, Long_2018}.

Indirect evidence of planet formation has been also found from observations of molecular lines at high angular and spectral resolution. Protoplanetary disks rotate around the central star following a sub-Keplerian speed unless perturbed by embedded planets or external dynamical interactions. 
The perturbation of CO kinematics in the disks around HD 163296 \citep[][]{Pinte_2018, Teague_2018}, HD 100546 \citep{Casassus_2019}, HD 97048 \citep{Pinte_2019}, and around eight of the DSHARP disks \citep{Pinte_2020}, has been interpreted as observational signatures from embedded planets in these disks. In all cases, the observed structures in channel maps and velocity moments align with predictions of kinematic signatures resulting from perturbing planets \citep[e.g.,][]{Perez_2015, Perez_2018, Izquierdo_2021}.

To date, the disk around PDS 70 \citep{Isella_2019b, Benisty_2021} is the best laboratory system to study planet disk interactions. It stands out as the only system where confirmed direct evidence of protoplanets has been found. Recent observational evidence from circumplanetary disks (CPDs) has emerged through high-angular-resolution and sensitive observations of this system \citep{Keppler_2018, Haffert_2019, Isella_2019b, Benisty_2021}. These observations revealed compact dust continuum emission around PDS 70c, located within a deep gap at 34 au from the disk center. 

A CPD/protoplanet candidate has been also proposed using molecular line emission at ALMA wavelengths in the disk around AS 209 and Elias 2-24 \citep{Bae_2022, Galloway_2023, Pinte_2023}, and using the Subaru Telescope and Hubble Space Telescope in AB Aur \citep{Currie_2022}.
The ages of PDS 70, AS 209, Elias 2-24, and AB Aur are estimated to be only a few Myr, suggesting that massive planets have already formed in some disks (at least) at the Class II stage. Additional observational evidence of planets embedded on protoplanetary disks is needed to understand how and when planet formation starts, and how the expected fast radial migration of solids can be prevented.

In this paper, we unveil hints of planet formation signatures in the large-cavity disk around 2MASS J16120668-3010270 (J16120 hereafter). The disk was observed with ALMA at Band 6, as part of the ALMA Large Program ``AGE-PRO: ALMA survey of Gas Evolution in Protoplanetary disks'' (Zhang et al. in prep.). The observations encompass dust continuum emission and four detected molecular lines: $^{12}$CO (J=2-1), $^{13}$CO (J=2-1), C$^{18}$O (J=2-1), H$_2$CO (J$=3_{0,3}-2_{0,2}$).

J16120 is a M0 star with an estimated age between 5-10 Myrs, a luminosity of $0.25 \ L_{\odot}$, a radius of $1.2 R_\odot$, a mass accretion rate of $4 \times 10^{-10} \ M_\odot \ \rm year^{-1}$,  \citep{Fang_2023}, and a mass of $0.7 M_\odot$ (this work). This star is localized in the Upper Scorpius star forming region, at a distance of 132.1 pc \citep{Gaia_2023}. It is situated far from most of the O and B stars in the Upper Scorpius star-forming region, distinguishing it as the disk with the lowest external far-ultraviolet irradiation field among the AGE-PRO sample in Upper Scorpius (Anania et al. in prep.). The morphology of the disk around J16120 has not been studied to date, even though it was detected and classified as a protoplanetary disk, encompassing evolved, transitional, and full disks by \cite{Fang_2023}.

This paper follows the structure outlined below: Section \ref{sec:observations} provides a summary of the ALMA Band 6 dust continuum observations and the detected molecular lines. Our analysis of the dust continuum visibilities and the disk morphology is detailed in Section \ref{sec:continuum}. Sections \ref{sec:lineCO} and \ref{sec:line_isotop} focus on the study of molecular line kinematics and moment maps. The discussion of results and observed disk substructures is found in Section \ref{sec:discussion}, followed by the paper's summary and conclusions in Section \ref{sec:conclusions}.

\begin{figure*}
    \centering
    \includegraphics[width=\textwidth]{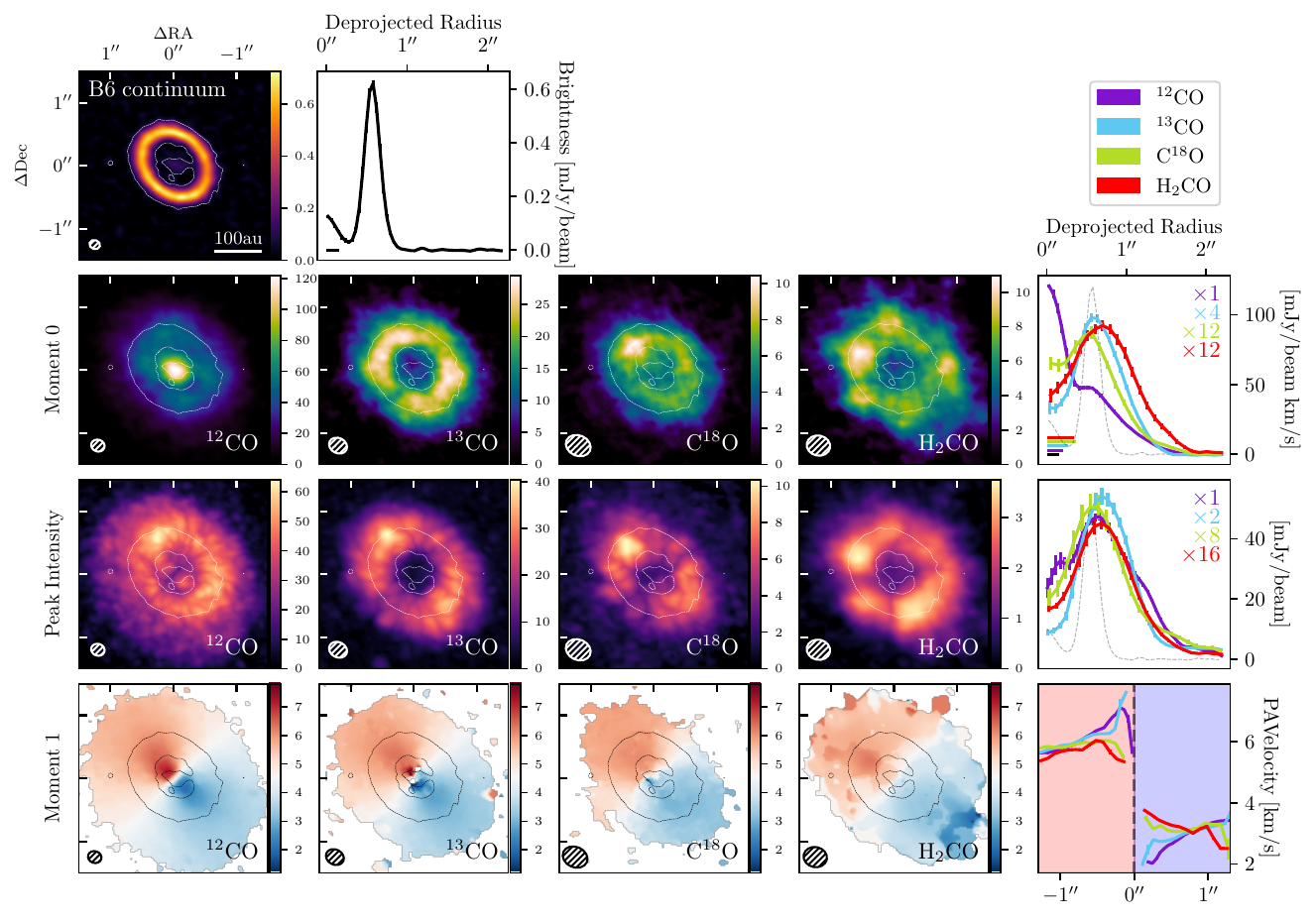}
    \caption{Continuum and molecular emission of the disk around J16120. First row: Dust continuum map (left) and dust continuum radial profile (right). 
    Second row from left to right: $\rm ^{12}CO \ (J=2-1)$, $\rm ^{13}CO \ (J=2-1)$, $\rm C^{18}O \ (J=2-1)$, and $\rm H_2CO \ (J=3_{(0,3)}-2_{(0,2)})$ moment 0 maps and azimutally average radial profiles.
    Third row: peak intensity maps for the molecular lines in the above panels in addition to the continuum radial profile.
    Fourth row: Moment 1 maps for the molecular lines in the above panels and velocity along the disk major axis. The beam is shown as a dashed ellipse in the bottom left corner in each panel, and in the bottom left corner of the moment 0 map radial profile panel. Dust continuum iso-contour at 3$\sigma$ is shown in all panels.}
    \label{fig:observations}
\end{figure*}

\begin{table*}
    \centering
    \caption{Image parameters}    
    \begin{tabular}{c|ccccc}
    \hline \hline 
    \multirow{2}{*}{Data}  & \multirow{2}{*}{Robust} & Beam & rms$^{*}$ & Vel resolution \\
    & & (mas $\times$ mas; deg) & (mJy beam$^{-1}$) &(km s$^{-1}$) & \\
    
    \hline
    Continuum & 0.0 &  171 $\times$ 142; 71.9 & 0.021 & - \\
    $^{12}$CO (J=2-1) & 0.0 & 217 $\times$ 188; 71.2 & 2.51 & 0.1 \\
    $^{13}$CO (J=2-1) & 0.5 & 297 $\times$ 241; 62.8 & 1.22 &0.2\\
    C$^{18}$O (J=2-1) & 1.0 & 407 $\times$ 324; 66.1 & 0.56 &0.2\\
    H$_{2}$CO ($J=3_{0,3}-2_{0,2}$) & 1.0 & 347 $\times$ 319; 78.4 & 0.14 & 1.6 \\
    \hline \hline
    \end{tabular}
    \newline
    $^{*}$ For line maps, this is the rms per channel.
    \label{tab:Imaging}
\end{table*}

\section{Observations}\label{sec:observations}

The ALMA Band 6 data in this work was obtained as part of the ALMA Large Program ``AGE-PRO: ALMA survey of Gas Evolution in Protoplanetary disks" (ALMA Project ID: 2021.1.00128.L),  during March and July 2022. 
The observations of J16120 consist in three short baseline (SB) executions and five long baseline (LB) executions. The execution block UIDs uid://a002/xf6d51d/x931e was not included in the final dataset due to semi-passed quality insurance. Each correlator setup consists of 6 spectral windows: one for continuum observations with a velocity resolution of 1.4 km s$^{-1}$, and five for molecular line emission, with a velocity resolution of 0.09 km s$^{-1}$. The SB and LB executions were self-calibrated using CASA version 6.4.3.27.
Continuum-only datasets were obtained by flagging the molecular line emission in each spectral window, and averaging to a maximum 125 MHz channel width. For astrometric alignment, we created low angular resolution continuum images of each dataset and fit a 2D gaussian to identify the disk center of each execution. Then, we align them to a common phase center (J2000 16h12m06.664505s -30d10m27.61789s).

Self-calibration was first performed using the SB data only. Before concatenating the short baselines, we checked for flux-scale alignment using the deprojected visibilities of each SB execution, and all of them have a similar total flux.
During the first self-calibration step, all the spectral windows and scans were combined to improve the signal-to-noise ratio (SNR) using \texttt{combine=`spws,scans'} within the \texttt{gaincal} task, and the initial solution time interval was set to \texttt{solint=`inf'}. We use the task \texttt{applymode=`calonly'} to calibrate data only and do not apply flags from solutions. The SNR improved by 47\% after the initial self-calibration step. No further self-calibrations were applied to the SB datasets because the SNR did not significantly increase for smaller solution time intervals.

Before concatenating the self-calibrated SB data with the LB executions, we double-checked for flux-scale alignment and found that the total flux in LB executions was 14\% higher than that in the self-calibrated SB. The SB executions were observed closer in time to the amplitude measurements of the flux calibrator from the ALMA calibrator monitoring\footnote{\url{https://almascience.eso.org/sc/}}. Therefore, the LB visibilities were re-scaled to match the SB flux.

After concatenating the self-calibrated SB and the re-scaled LB datasets, we self-calibrated the data following the same strategies and parameters previously described. We did not apply amplitude self-calibration since we did not observe significant SNR improvement in this case. 
To obtain the molecular line data sets, we apply the astrometric alignments, flux-scale alignment, and resulting calibration tables 
to the original datasets that had no spectral averaging. Then, we used the task \texttt{uvcontsub} to extract the continuum emission from the dataset, and finally we split out frequency ranges corresponding to the different molecular lines.
Full comprehensive details regarding the observational setup and self-calibration process for all the disks in the Upper Scorpius star-forming region will be provided in Agurto-Gangas et al. in prep.

In order to study the small gas and dust substructures in the disk around J16120, we prioritize increasing the angular resolution of the dust continuum and detected molecular lines observations ($\rm ^{12}CO$ (J=2-1), $\rm ^{13}CO$ (J=2-1), $\rm C^{18}O$ (J=2-1), and $\rm H_2CO$ ($\mathrm{J=3_{(0,3)}-2_{(0,2)}}$)). Therefore, in this study, we re-image the data by adopting a smaller Briggs robust parameter, aiming for higher resolution compared to Agurto-Gangas et al. in prep., where all the disk observations in Upper Scorpius were imaged at the same angular resolution ($\sim 0.27$ arcsec) for comparison. The choice of the Briggs robust parameter for each molecular line map is optimized to strike a balance between angular resolution and signal-to-noise ratio (SNR). In particular, we choose a robust of 0.0, 0.5, 1.0, and 1.0 for imaging the $^{12}$CO, $^{13}$CO, C$^{18}$O, and H$_2$CO molecular emission, respectively, and 0.0 for the dust continuum map.

We use the \texttt{tclean} task in CASA, with a multi-scale algorithm corresponding to point sources and scales of half, one, two, and three times the beam Full Width at Half Maximum (FWHM). We adopted a clean threshold of 1$\sigma$ and 4$\sigma$ for the dust continuum and molecular lines, respectively. The 4$\sigma$ criteria was chosen following the same imaging strategies in the MAPS ALMA Large Program \citep{Oberg_2021, Czekala_2021}. In addition, we applied the JvM correction to the molecular emission maps \citep[][]{Jorsater_1995, Czekala_2021}, which corrects the units of the residual maps from Jansky per dirty beam to Jansky per CLEAN beam. We obtain a residual correcting factor of $\epsilon = 0.71, 0.47, 0.33$, and 0.31 for $\rm ^{12}CO, ^{13}CO, C^{18}O$, and $\rm H_2CO$ respectively. Table \ref{tab:Imaging} shows the final imaging properties for each molecular line and dust continuum emission.
We double-check that the observed disk sub-structures in the following sections are not dependent on the inclusion of the JvM correction (Appendix \ref{app:Supporting-Figs}). In addition, we also computed all the results involving molecular line emission in Sections \ref{sec:lineCO}-\ref{sec:line_isotop} without including the JvM correction, and we confirm that the final results do not depend on this correction. The standardized flux measurements for each molecular line for the entire Upper Scorpius region in AGE-PRO (including J16120) will be presented in Agurto-Gangas et al. in prep.

\begin{figure*}
\centering    
    \includegraphics{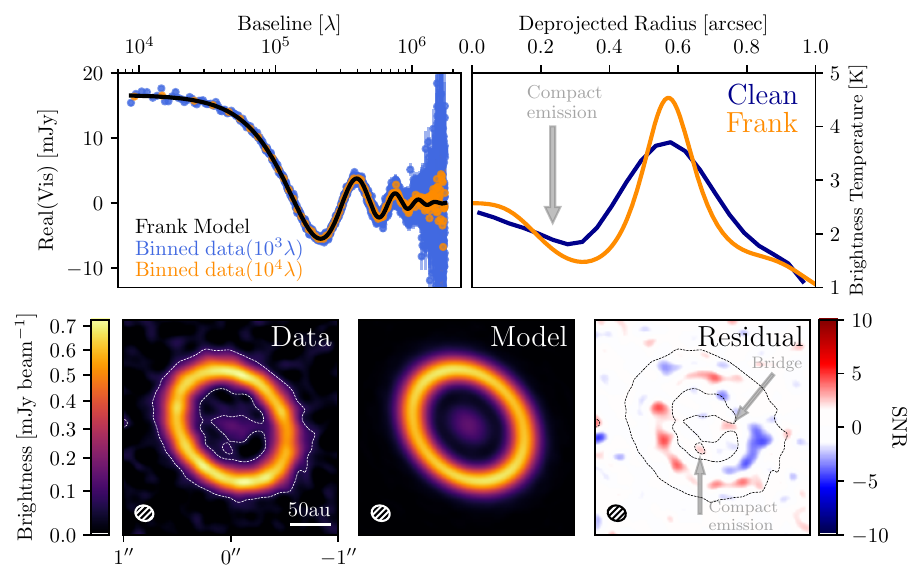}
    \caption{Dust continuum modeling of the disk around J16120. Top left panel: Observed binned visibilities (blue and orange dots, at a bin size of $10^{3}$ and $10^{4} \lambda$, respectively), and visibility model (black curve) of the dust continuum data. Top right panel: Deprojected continuum radial profile from the image plane (blue curve) and from the visibility fit non-convolved model (orange curve). The arrow marks the radial position of the compact emission. Bottom left panel: \textsc{CLEAN} image of the dust continuum data. Bottom middle panel: \textsc{CLEAN} image of the visibility model. Bottom right panel: \textsc{CLEAN} image of the visibility residuals. The iso-contour in the bottom panels is at a 3$\sigma$ level.}
    \label{fig:continuum-analysis}
\end{figure*}

\section{Results}\label{sec:results}
\subsection{Continuum and molecular moment maps}\label{sec:maps}
The continuum and molecular emission maps of J16120 are shown in Figure \ref{fig:observations}. The top panels show the dust continuum emission map and radial profile. The second, third and fourth row show the moment 0 maps (integrated intensity over the spectral line), peak intensity maps, and moment 1 maps (integrated intensity-weighted velocity over the spectral line), respectively, of four detected lines: $\rm ^{12}CO$ (J=2-1), $\rm ^{13}CO$ (J=2-1), $\rm C^{18}O$ (J=2-1), and $\rm H_2CO$ ($\mathrm{J=3_{(0,3)}-2_{(0,2)}}$), from left to right. 
The deprojected radial profiles are shown in the right panels of each row (the disk inclination and position angle are computed in Section \ref{sec:continuum}). The bottom right panel shows the velocity in a 5-degree slice around the disk's major axis.

The moment maps were computed using the \textsc{bettermoments} tools described in \cite{Teague_2018b, Teague_2019}. We include a Keplerian mask using the disk geometry (computed from the visibilities, see next section for details), a systemic velocity of 4.52 km s$^{-1}$, a central mass of 0.72 $M_{\odot}$, an emitting surface given by $z(r)$ $= 0.17^{\prime \prime} \times (r/1^{\prime \prime})^2$,  and 2 beam smearing. The data are clipped using a threshold of 1$\sigma$ and 3$\sigma$ for the moment 0 and moment 1 maps, respectively.

The disk shows a ring structure in the dust continuum and $\rm ^{13}CO$, $\rm C^{18}O$, and $\rm H_2CO$ moment 0 maps. The $\rm ^{12}CO$ also shows a ring structure, but it overlaps with the emission from the inner disk. The radial extent of the dust continuum emission is smaller than that from the molecular emission, similar to most of the known resolved protoplanetary disks \citep[e.g.,][]{Andrews_2012, Ansdell_2018, Law_2020_rad, Long_2022b}, and as expected from dust evolution models \citep[e.g.,][]{Trapman_2019, Birnstiel_2023}.

We double-checked that all our imaging strategies are not creating artificial moment 0 structures. The non axi-symmetric structures in the moment 0 maps of $^{13}$CO and C$^{18}$O are below 5$\sigma$, except for the bright emission in the northeast of C$^{18}$O, where the excess emission is $\sim 6 \sigma$, and could be tracing a higher temperature or gas density. This bright emission is also observed in the peak intensity maps of the three CO isotopologues, although they are slightly spatially shifted, likely due to differences in emission heights or spatial smearing effects caused by differences in the angular resolution. In all cases, the emission in this region is $\gtrsim 5 \sigma$ when the averaged azimuthal intensity is subtracted.

On the other side, the velocity resolution of the $\rm H_2CO$ data set (Table \ref{tab:Imaging}) only allow us to detect this molecule in three independent channels. Therefore, its moment 0 map looks like a hexagon (2 surfaces per channel). 

We model the visibilities of the dust continuum emission and we study its azimuthal asymmetries in Section \ref{sec:continuum}. The moment 1 maps, which exhibit a rotation around the central protostar, are employed in Sections \ref{sec:lineCO} and \ref{sec:line_isotop} to constrain the stellar mass, the $^{12}$CO emission surface, and the gas kinematics.

\subsection{Dust continuum emission morphology}\label{sec:continuum}

We model the visibilities of the dust continuum emission of J16120{\footnote{The visibility analysis of the 30 protoplanetary disks in the AGE-PRO ALMA Large Program is addressed in Vioque et al. in prep.}} with \textsc{Frankenstein} \citep{Jennings_2020}, a \textsc{Python} library that fits the visibilities using a non-parametric radial brightness profile, assuming azimuthal symmetry. The interferometric visibilities are extracted from the self-calibrated measurement sets using a modified version of the \textit{export\_uvtable} function in \cite{uvplot_tazzari}. In our version, the uv-distances are normalized using the wavelength of each channel and spectral window, instead of the average wavelength of the whole data set. Beyond the emission of the dust continuum disk, a very faint additional source is detected (see a larger continuum map in Appendix \ref{app:Supporting-Figs}) at a projected distance of $\sim 4.5^{\prime \prime}$ ($\sim 600$ au at the J16120 distance) away from the disk center. While we mask and subtract the visibilities of the additional source, its presence has negligible effects on the final visibility modeling due to its considerable faintness and the substantial distance from the disk.

The disk offset ($\Delta \rm RA, \Delta \rm Dec$) with respect to the phase center is computed from minimizing the imaginary part of the visibilities \citep[e.g.,][]{Isella_2019}. The disk inclination ($i$), and disk position angle (PA, measured counterclockwise from North) are computed from minimizing the spread of the real part of the visibilities \citep[e.g.,][]{Isella_2019}. We used a Markov chain Monte Carlo (MCMC) method \citep[implemented in the \textsc{Python} library \textsc{emcee},][]{Foreman_2013} to explore the space parameter of the offset and geometry. The starting point of the Markov chain is chosen based on an initial guess of the geometry and offset in the image plane, and uniform priors were included for all parameter. The inferred disk offset (relative to the phase center in Section \ref{sec:observations}) and geometry are $\Delta \rm RA = 4.6 ^{+20}_{-20}$ mas, $\Delta \rm Dec = 6.6 ^{+10}_{-10}$ mas, $\rm inc= 37.0 ^{+0.1}_{-0.2} $ deg, $\rm PA = 45.1^{+0.2}_{-0.9} $ deg.

The \textsc{Frankenstein} reconstructed radial profile depends on two hyper-parameters: $\alpha$, $w_{\rm smooth}$. The former mimics the SNR threshold where the data are not taken into account when fitting the visibilities, and the latter controls the smoothness of the fit to the power-spectrum (see full details in \cite{Jennings_2020} and \cite{Jennings_2021}). We tested a grid of hyper-parameters within $1.05 \leq \alpha \leq 1.3$ and $10^{-4} \leq w_{\rm smooth} \leq 10^{-1}$, recommended by \cite{Jennings_2020}, and found no important differences in the reconstructed brightness radial profile. Therefore, we fix them to $\alpha = 1.05$ and $w_{\rm smooth} = 10^{-4}$.

The top left panel in Figure \ref{fig:continuum-analysis} shows the real part of the observed visibilities (binned at $10^{3}\lambda$ and $10^{4} \lambda$) as a function of baseline, and the visibility model obtained with \textsc{Frankenstein}. The model is a good representation of the observed visibilities, and it  is able to find a visibility sub-structure at long baselines ($\gtrsim 10^6 \lambda$). The top right panel shows the deconvolved brightness temperature from the visibility fit (orange line), and the azimuthally averaged brightness temperature computed from the \textsc{clean} image of the data (blue line).  Both profiles show a ring structure with a peak position at 0.57$^{\prime \prime}$ (75 au), and an inner disk. Note that the width of the main ring obtained with \textsc{Frankenstein} is narrower than that from \textsc{clean}. This difference occurs because the radial profile obtained with the former is deconvolved, while the latter is limited by the angular resolution in the image plane.

The gray arrow marks the deprojected radial distance of a compact dust continuum emission, which can be seen in the dust continuum map in the bottom left panel. It lies between the disk ring and the inner disk, at a
projected radius of 0.20$^{\prime \prime}$ (26.4 au) and deprojected radius of 0.23$^{\prime \prime}$ (30.4 au), and at an azimuthal position (with respect to the North-East major axis) of $\sim$126 deg, in the deepest region of the disk gap. In addition, there is a tentative emission bridge connecting the Western part of the inner disk and the ring. We discuss possible origins of these substructures in Sections \ref{sec:Discussion_continuum} - \ref{sec:CPD}.

Note that the inclination and position angle of the inner disk looks different compared with that from the main ring (bottom left panel of Figure \ref{fig:continuum-analysis}). However, due to the low SNR in the inner disk, the geometry from this region is poorly constrained, and we assume the same geometry in all the disk.

\begin{figure*}
    \centering
    \includegraphics[scale=0.8]{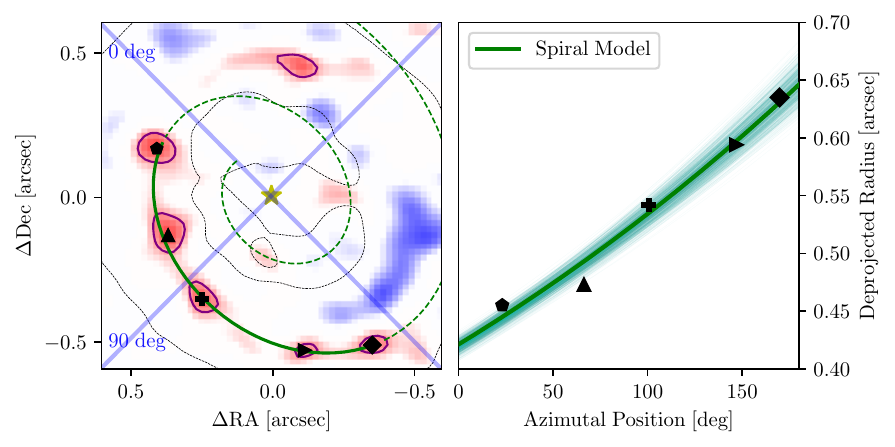}
    \caption{Spiral arm structure inferred from the disk azimuthal asymmetries. Left panel: Zoom into the residual map in Figure \ref{fig:continuum-analysis}. Black markers are the position of the spiral arm structure, the green solid line is the best fit spiral model, and its extrapolation is shown as a green dashed line. The purple lines around the red knots are 3$\sigma$ iso-contours. 
    The blue lines at 0 and 90 degrees are the disk major and minor axis, respectively.
    Right panel: Azimuthal position and deprojected radius of the spiral arm structure (black markers), and best fit model (green solid line). The thin turquoise solid lines represent 1000 random chains from the posterior distribution.}
    \label{fig:continuum-residual}
\end{figure*}

The clean image of the disk model is shown in the bottom middle panel of Figure \ref{fig:continuum-analysis}. The model is computed using the deconvolved radial profile from \textsc{Frankenstein} and sampling its visibilities at the same observational uv-coverage. This model is a good representation of the axisymmetric morphology of the dust continuum map. The azimuthal asymmetries of the data with respect to the axisymmetric model are shown in the bottom right panel (The residuals were computed in the visibility space, and then the visibility residuals were imaged with the same parameter as the original image). Note that the residual map shows negative values (in blue) in the South-West of the ring, where the ring is faint compared with the average azimuthal intensity. There are also positive residuals (in red) at the position of the compact emission, the bridge, and within the ring. The latter follows spiral arm structure, as shown in Figure \ref{fig:continuum-residual}. 

The residual map can be affected by errors in the disk geometry or the offset of the disk center with respect to the phase center \citep{Andrews_2021}. We explored these effects in Appendix \ref{app:Geom-Residuals}, where we show that the offset can significantly modify the spiral arm structure, while the disk geometry has a lower impact on the residuals. However, our offset and geometry estimations minimize the residual structures. We also show that the residual structures can be observed directly in the clean image of the data (model independent) when the color bar is adjusted to highlight the bright emission of the disk, confirming the azimuthal asymmetries in the residual map. On the other hand, the residuals corresponding to the compact emission and the bridge are always observed, independent of the geometry and offset constraints within our error bars.

We analyze the spiral arm structure by using the deprojected radius and azimuthal position as references (with respect to the disk major axis in the North-East), as marked in the left panel of Figure \ref{fig:continuum-residual}. These coordinates correspond to the brightest pixel within each red knot. The deprojected coordinates are then fitted using the logarithmic spiral arm prescription outlined in \cite{Perez_2016},

\begin{equation}
    r(\theta) = R_0 \exp(b\theta),
\end{equation}
where $R_0$ is the radius at the azimuthal position $\theta=0$, and $b$ is the rate at which the spiral increases their distance from the origin. We found $R_0 = 0.42^{\prime \prime}\pm 0.01 ^{\prime \prime}$, $b = (2.4 \pm 0.2) \times 10^{-3} \ \rm deg^{-1}$. The right panel in Figure \ref{fig:continuum-residual} shows the best fit (green dashed line) to the spiral arm structure and 1000 random chains from the posterior distribution. The best fit is extrapolated (green solid line) and plotted in the left panel. The extrapolated curve matches with the position of the compact emission and the bridge. The only remaining red residual is located at the North of the disk. The extrapolation of the spiral arm does not match the position of this knot, and its origin is not clear. Possible explanations to the origin of the dust continuum sub-structures are discussed in Section \ref{sec:discussion}.

\begin{figure*}
    \centering
    \includegraphics[width=1.0\textwidth]{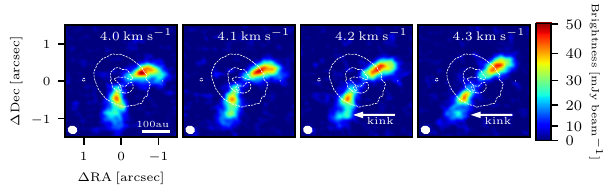}
    \caption{Selected $^{12}$CO channel maps of J16120. The white iso-contour is taken from the dust continuum emission in Figure \ref{fig:continuum-analysis}. The synthesized beam and the channel velocity are shown in the bottom-left and top-right corner of each panel, respectively. The color bar is not linear but slightly adjusted to highlight the kink emission.}
    \label{fig:CO-channels}
\end{figure*}


\subsection{$^{12}$CO kinematic properties}\label{sec:lineCO}

The observed $^{12}$CO channels of J16120 reveals interesting kinematic signatures. Figure \ref{fig:CO-channels} shows selected $^{12}$CO channel maps where a perturbation is observed in the South of the disk, beyond the dust continuum ring \citep[similar to AS 209,][]{Bae_2022}, and at a deprojected radius of $\sim 0.85^{\prime \prime}$ (112 au). This kind of kinematic perturbation, as discussed in Section \ref{sec:introduction}, is expected from the dynamical interaction between the disk and an embedded planet. Additional $^{12}$CO channels can be seen in Appendix \ref{app:Supporting-Figs}.

\begin{figure}
    \centering
    \includegraphics[scale=0.55]{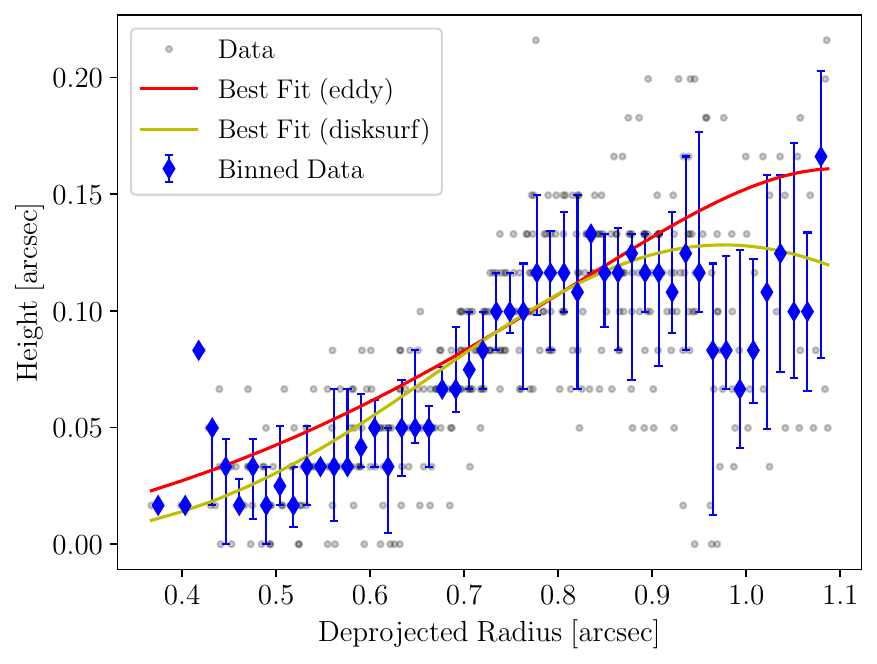}
    \caption{Emission surface extracted from the CO channel maps (grey dots), and binned surface (blue diamonds). The error bars are computed from the 16th and 84th percentiles.
    The surface parametrization from the velocity map and selected channel maps is shown as a red and yellow line, respectively. The emission surface measurements beyond 1.1$^{\prime \prime}$ have a large scatter due to low SNR.}
    \label{fig:CO-surface}
\end{figure}

{
The emission from CO usually does not trace the disk midplane \citep[e.g.,][]{Dutrey_2017, Law_2021, Paneque_2023}, but rather an emission surface, which is usually parameterized as
\begin{equation}
    z(r) = z_0 \left( \frac{r}{1^{\prime \prime}} \right)^{\psi} \exp\left( - \left[ \frac{r}{r_{\rm taper}}\right]^{q_{\rm taper}} \right),
    \label{eq:EmissionSurface}
\end{equation}
where $z_0$ is the emission surface at a reference radius, $r$ is the deprojected radius, $r_{\rm taper}$ is the taper radius, and the exponent power-laws $\psi$, $q_{\rm taper}$ modulates the slope of the emission surface in the inner and outer disk, respectively.

The free parameters of the emission surface (Equation \ref{eq:EmissionSurface}) can be estimated by extracting the coordinates of the emission surface from selected channels and fitting the extracted data \citep[e.g.,][]{Pinte_disksurf_2018, Law_2021}, or using the velocity field map of the disk and fitting a Keplerian model \citep[e.g.,][]{Teague_2021}. We use both methodologies and compare the results.

In the former, we use the Python package \textsc{disksurf} \citep{Pinte_disksurf_2018, disksurf} to extract the surface in different channels. We masked all the surface measurements that had negative values or had unrealistic aspect ratio $z/r > 1$. The unmasked surface measurements are fitted using the height parametrization (Equation \ref{eq:EmissionSurface}). The extracted surface in different channels and the emission surface parametrization are shown in Figure \ref{fig:CO-surface}. Note that the emission height goes to 0 at $r \approx 0.4^{\prime \prime}$, which is the radius where the dust continuum emission show a clear gap (Figure \ref{fig:continuum-analysis}).

We also fit the $^{12}$CO emission surface from the velocity field map using the Python package \textsc{Eddy} \citep{eddy}.}  In this modeling, the velocity map is fitted using a Keplerian velocity, taking into account the height above the midplane ($z$) as: $v_{\rm kep} = \sqrt{ GM_{*} r^2/(r^2+z^2)^{3/2}}$.
The velocity map is computed using the \texttt{quadratic} methodology described in \cite{Teague_2018b}, which is part of the \textsc{bettermoments} tools. This velocity map has been proven to be a robust method for measuring centroids of spectral lines \citep{Teague_2018b}. However, we also computed velocity maps using additional methodologies (including moment 1 maps), and similar constraints to the emission surface were found (see Appendix \ref{app:velocity}).

We use \textsc{Eddy} \citep{eddy} to explore the posterior probability distribution of the stellar parameters and emission surface. The disk inclination and position angle are fixed using the dust continuum constrains computed in the previous section. We fit the emission within a deprojected radius of $0.33^{\prime \prime}$ (1.5 times the beam major axis) and $1.4^{\prime \prime}$ (where the SNR $\lesssim 3$). The emission surface is also parameterized using Equation \ref{eq:EmissionSurface}.

The offset of the disk center ($\delta x_0$, $\delta y_0$), the mass of the central star ($M_*$), and the disk systemic velocity ($v_{\rm lsr}$) are also free parameters of the fit. A summary of the best fit parameters computed from the velocity map is given in Table \ref{tab:Surface}, and the corner plots of the posterior distribution of each free-parameter is shown in Appendix \ref{app:MoreKinematics}. The parametric emission surface computed from the velocity map with \textsc{eddy} is also shown in Figure \ref{fig:CO-surface}.

Both parametrizations (selected channels with \textsc{disksurf}, and velocity map with \textsc{eddy}) show consistency within the error bars of surface measurements (represented by the binned data). Disagreement between them only occurs beyond approximately $\sim 1^{\prime \prime}$, where the SNR decreases.

\begin{table*}
    \centering
    \caption{Best fit parameters computed from different molecular line emission. The emission surface is only fitted for $^{12}$CO.}
    \begin{tabular}{c|c|cccc}
        \hline \hline
               &      \textsc{disksurf} & \multicolumn{4}{c}{\textsc{eddy}} \\
         Parameter & $^{12}$CO& $^{12}$CO & $^{13}$CO & C$^{18}$O & H$_2$CO \\
         \hline
         $\delta x_0$ [mas]            & -35.0$^{*}$& -35.0$^{+1.0}_{-1.1}$ & -1.7$^{+0.2}_{-0.2}$ & -67.8$^{+0.6}_{-0.6}$ & -22.4$^{+0.1}_{-0.1}$ \\
         $\delta y_0$ [mas]            & -1.0$^{*}$ & -1.0$^{+1.0}_{-1.0}$  & -9.6$^{+0.2}_{-0.2}$ & 0.4$^{+0.6}_{-0.6}$   & -26.8$^{+0.1}_{-0.1}$ \\
         $M_*$        $[M_{\odot}]$    & 0.70$^{*}$ & 0.70$^{+0.1}_{-0.1}$  & 0.66$^{+0.1}_{-0.1}$ & 0.77$^{+0.1}_{-0.1}$  & 0.75$^{+0.1}_{-0.1}$   \\
         $v_{\rm lsr} ^{(a)}$ [km s$^{-1}$]   & 4.5$^{*}$ & 4.5$^{+0.1}_{-0.1}$   & 4.5$^{+0.1}_{-0.1}$  & 4.5$^{+0.1}_{-0.1}$   & 4.5$^{+0.1}_{-0.1}$   \\
         $z_0$ [arcsec]                & 0.58$^{+0.91}_{-0.30}$ & 0.17$^{+0.01}_{-0.01}$  & -    & -     & -     \\
         $\psi$                        & 3.9$^{+0.8}_{-1.3}$& 2.0$^{+0.1}_{-0.1}$   & -    & -     & -     \\
         $r_{\mathrm{taper}}$ [arcsec]  & 0.86$^{+0.23}_{-0.31}$ & 1.3$^{+0.2}_{-0.2}$   & -    & -     & -     \\
         $q_{\mathrm{taper}}$          & 2.6$^{+1.5}_{-1.3}$ & 8.3$^{+0.8}_{-0.7}$   & -    & -     & -     \\
         \hline \hline
    \end{tabular}
    \newline
    Error bars are computed from the $1\sigma$ posterior corner plots. However, a more realistic velocity uncertainty is given by the spectral resolution and angular resolution for each molecule in Table \ref{tab:Imaging}. $^{*}$Fixed parameters during the fitting. 
    \label{tab:Surface}
\end{table*}

\begin{figure*}
    \centering
    \includegraphics{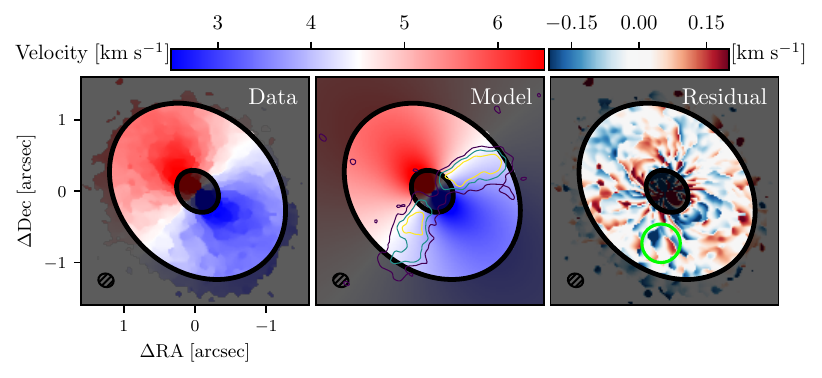}
    \caption{$^{12}$CO velocity modeling of the disk around J16120. Left: Observed velocity structure in $^{12}$CO. Middle: Best fit Keplerian model. Emission contours corresponding to 4$\sigma$, 8$\sigma$, and 12$\sigma$ of the 4.2 km s$^{-1}$ channel map (Figure \ref{fig:CO-channels}) are included as a spatial reference. Right: Velocity residuals . The green circle highlights a velocity structure at the position of the kinematic perturbation shown in Figure \ref{fig:CO-channels}.}
    \label{fig:CO-modeling}
\end{figure*}

Figure \ref{fig:CO-modeling} shows the $^{12}$CO velocity map (left panel), the best fit model using the emission surface parametrization from the velocity map (middle panel), and the velocity residuals (right panel). Appendix \ref{app:MoreKinematics} shows the velocity best fit model and velocity residuals when the surface emission parametrization is fixed to that obtained from $^{12}$CO channel maps with \textsc{disksurf}. There are not important differences between the two models.

Upon inspecting the velocity residuals in Figure \ref{fig:CO-modeling}, we highlight a residual structure within the green circle, close to the position where the kinematic signature is observed in the channel maps (Figure \ref{fig:CO-channels}). There are negative and positive residuals in the North-West and South-East of this structure, respectively, with a smooth transition between them.  The inclination and position angle are  similar to that of the dust continuum emission. This structure is similar to the expected rotation of circumplanetary disks \citep[e.g.,][]{Perez_2015, Izquierdo_2021}. Possible origins of the kink and velocity residuals are discussed in Section \ref{sec:Discussion_CO}. 

\begin{figure*} 
    \centering
    \includegraphics[width=\textwidth]{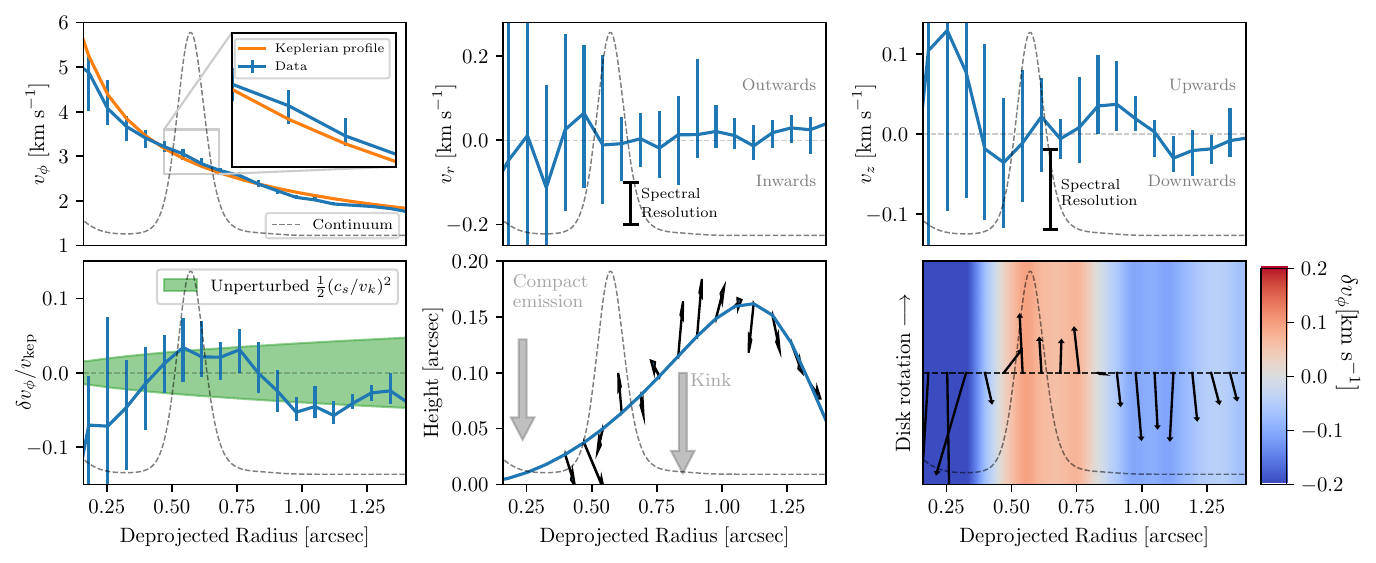}
    \caption{Gas kinematics inferred from the $^{12}$CO channel maps in the disk around J16120. Solid blue lines in the top panels show the azimuthal, radial, and vertical velocity (from left to right) as a function of radius. Error bars are 1$\sigma$ uncertainties and the vertical bar in the middle and right panel shows the spectral resolution of the data. The orange line in the top-left panel is the Keplerian model in the mid-plane. The embedded plot in the top-left panel zooms in on the radial profiles around the dust ring.  
    Bottom left: Normalized deviation of the azimuthal velocity with respect to the Keplerian model (blue line), and expected velocity deviation of an unperturbed smooth disk (green shaded area). Bottom middle: Velocity field in the radius-height plane. The blue line is the parameterized $^{12}$CO emission surface. Bottom right: Velocity field in the radius-azimuth plane. The color bar shows the deviation of the azimuthal velocity with respect to the Keplerian model. The error bars of the velocity vectors in each direction can be seen in the radial profile panels.
    In all panels, the continuum radial profile is shown as a dashed gray line for reference.}
    \label{fig:CO-kinematics}
\end{figure*}

We also use \textsc{Eddy} and the $^{12}$CO channel maps to compute the deprojected velocity along the disk radius ($v_r$), azimuth ($v_\phi$), and vertical disk axis ($v_z$). We used the disk geometry, offset, systemic velocity, and emission surface previously computed to fit the $^{12}$CO velocity field. 
The top panels in Figure \ref{fig:CO-kinematics} show the three velocity components as a function of radius.
The azimuthal velocity ($v_{\phi}$, solid blue line) decreases with radius, and it is similar to the Keplerian velocity model in the mid-plane (solid orange line), which is computed using the stellar mass estimated from the $^{12}$CO emission line (see Table \ref{tab:Surface}.) The embedded plot zooms in on the Keplerian profile and azimuthal velocity around the dust continuum ring, where the difference between both profiles are on the order of the azimuthal velocity error bars.
The radial velocity oscillates around zero, and there is not a clear tendency due to the large error bars. 
The vertical velocity also has large error bars, especially at $\lesssim 0.6^{\prime \prime}$. However, $v_z$ tends to be positive (Upwards) between $\sim 0.75^{\prime \prime}$ and $1.0^{\prime \prime}$. Beyond this range (where the disk surface decays with radius), $v_z$ is negative (Downwards) and the $^{12}$CO is presumably moving towards the mid-plane.
Given the large error bars and inferred subsonic velocities ($c_s \sim 0.3$ km s$^{-1}$ for a gas with a temperature of 20K), it is difficult to identify the radial and vertical velocities as a robust signal of coherent motion.

The bottom left panel shows the normalized deviation of the azimuthal velocity with respect to the Keplerian velocity in the midplane, given by $\delta v_{\phi}/v_{\rm kep} = (v_{\phi} - v_{\rm kep})/v_{\rm kep}$. We plot the area below $\frac{1}{2} (c_s/v_{\rm kep})^2$, where $c_s$ is the sound speed computed from a passively irradiated disk and a 0.25 $L_{\odot}$ star \citep{Fang_2023}. This term represents the expected azimuthal velocity deviation (with respect to the Keplerian rotation) for a disk with no pressure bumps, but a smooth and monotonically decreasing pressure gradient (created by a smooth temperature and gas density radial profiles, responsible of dust radial drift), and can be used to investigate the effect of azimuthal deviations as pressure maxima, as discussed in Section \ref{sec:Discussion_continuum}.
Note that the azimuthal disk velocity tends to be super-Keplerian in the region around the dust continuum ring ($\delta v_{\phi} > 0$). However, due to the large error bars, it remains consistent with a Keplerian model. 

The velocity fields in the radius - height/radius - azimuth plane are shown in the bottom middle/right panel, respectively. The blue line in the former is the parameterized emission surface with \textsc{eddy}, and the gray arrows show the deprojected radial position of the compact emission and the kink for reference. The color bar in the right panel shows the central value of azimuthal velocity deviation. However, as discussed previously, the error bars are on the same order of magnitude.

\subsection{$^{13}$CO, C$^{18}$O and H$_{2}$CO kinematics}\label{sec:line_isotop}
The $^{13}$CO, C$^{18}$O and H$_{2}$CO molecular emission is faint compared with that of $^{12}$CO. Detecting faint emission from our data with a good SNR requires a higher robust value (lower resolution) compared with that used for $^{12}$CO. The synthesized beam sizes for $^{13}$CO, C$^{18}$O and H$_{2}$CO (Table \ref{tab:Imaging}) are a factor of 1.8, 3.2, and 2.6 larger compared with the $^{12}$CO synthesized beam, respectively. 
The $^{12}$CO velocity structure was fitted over a radial range $0.3^{\prime \prime}-1.4^{\prime \prime}$. For the rest of the molecular lines, this interval is only 3.6, 2.6 and 3.1 times the beam size for the $^{13}$CO, C$^{18}$O and H$_{2}$CO observations, respectively. This somewhat limits our ability to sample this region of the disk with precision in these tracers. 

Despite this limitation, we fit the velocity field for $^{13}$CO, C$^{18}$O and H$_{2}$CO using a flat disk (no emission surface is fitted), and we obtained a stellar mass and systemic velocity similar to that obtained from $^{12}$CO. The best fit parameters for all lines are summarized in Table \ref{tab:Surface}, and the velocity best fit models and residuals in Appendix \ref{app:MoreKinematics}. No significant residual structures are found in any of the faint molecular transitions at this resolution and sensitivity.

\section{Discussion}\label{sec:discussion}
\subsection{Possible origins of the dust ring} \label{sec:Discussion_continuum}

One of the most common origins of ring structures at millimeter wavelengths is dust traps. Dust traps mainly affect the spatial distribution of large grains \citep[millimeter to centimeter sizes, ][]{Pinilla_2012}, and thus, their signatures can be inferred from millimeter observations. If dust grains are trapped in the ring of J16120, the width of the observed ring should decrease with wavelength \citep[e.g.,][]{Pinilla_2012b, Sierra_2019}. The dust trap hypothesis can also be studied by inferring the grain size as a function of radius around the ring. This can be achieved by fitting the spectral energy distribution \cite[e.g.,][]{Carrasco-Gonzalez_2019, Sierra_2021}, and looking for evidence of local maxima in the grain sizes at the ring peak.

Dust traps originate from pressure bumps where the gas velocity changes from sub-Keplerian to super-Keplerian around the pressure peak. Consequently, dust traps can also be studied using $^{12}$CO kinematics. As mentioned in Section \ref{sec:lineCO}, the average azimuthal velocity over the dust continuum ring tends to be super-Keplerian, as shown in the embedded plot in the top-left panel of Figure \ref{fig:CO-kinematics}. 
However, the deviation of the azimuthal velocity from the Keplerian model in this work is not significant enough to definitively confirm or rule out any dust trap scenario. Note that the Keplerian profile falls within the error bars of the azimuthal velocity. Furthermore, the magnitude of the inferred azimuthal velocity deviation is low compared to the expected azimuthal deviation for smooth disks with no local pressure bumps, which is of the order of $\partial v_{\phi} \sim \frac{1}{2} c_s^2/v_{\rm kep}$ \citep{Takeuchi_2002, Rosotti_2020}. This value is shown as green shaded area in the bottom left panel in Figure \ref{fig:CO-kinematics}. The inferred velocity deviation around the dust continuum ring is only marginally below the unperturbed threshold, and it is also consistent with zero within the error bars. 

The location of a pressure maximum should coincide with a negative slope in the azimuthal velocity deviation and be greater in absolute value than that of an unperturbed pressure profile \citep{Rosotti_2020}. For J16120, a slightly small negative slope around the ring is inferred, but is also consistent with zero and positive slopes within the error bars. The central value of the slope between two consecutive radii (separated by $\Delta r = 72.4$ mas) is  $m \sim -0.15$ arcsec$^{-1}$. Thus, the total variation induced by tentative pressure bump is $|m| \Delta r \sim 1\%$, which is below the $\sim 2\%$ unperturbed value at the ring location. Given the uncertainties, we cannot rule-out or confirm a dust trap signature from these estimates.

Finally, errors in the estimation of the mass of the central star (e.g., inferred from different line isotopologues, see Table \ref{tab:Surface}) influence the region where sub-Keplerian or super-Keplerian velocities are inferred \citep{Rosotti_2020}, making it challenging to provide strong evidence for the dust-trapping mechanism based on our $^{12}$CO observations.

On the other side, the central values of the radial velocity shows a flip between gas moving radially outwards and radially inward around the peak of the dust continuum ring (right middle panel, Figure \ref{fig:CO-kinematics}). However, given the large uncertainty in the gas radial velocity and the low subsonic velocities (left middle panel, Figure \ref{fig:CO-kinematics}), it is difficult to know if gas is dragging dust grains towards the position of the dust continuum ring by dynamical coupling, which could explain its origin.

Dust trap signatures from both continuum and gas kinematics require constraints on the radial profile of the dust grain size. Due to the lack of high angular resolution observations at different wavelengths, we cannot address this problem. However, we can estimate the radially averaged grain size using the averaged spectral index between ALMA Band 6 ($226$ GHz) and Band 7 ($341$ GHz) using the integrated flux density constraints in this work and in Carpenter et al. in prep., where J16120 was observed at ALMA Band 7 with an angular resolution $\sim 0.25$ arcsec, but only a few minutes on source. The integrated flux density at Bands 6 and 7 is 16.6 and 65.5 mJy, respectively. Thus, the average spectral index is $\alpha = 3.3 \pm 0.3$, consistent with the expected spectral index for a 75 au cavity radius \citep{Pinilla_2014}, and the estimated spectral index between 1.3 and 1.05 mm from AGE-PRO (Deng et al. in prep., Kurtovic et al. in prep.).
This spectral index typically traces optically thin emission. In this regime, the opacity spectral index $\beta$ can be estimated as $\beta = \alpha - 2 = 1.3 \pm 0.3$. This corresponds to maximum grain sizes of 1.6 mm and 4.0 mm for particle size distribution $n(a)da\propto a^{-p}da$ (number of particles with a size between $a$ and $a+da$) with $p=2.5$ and $p=3.5$, respectively, based on the DSHARP opacities \citep{Birnstiel_2018}.

\begin{figure}
    \centering
    \includegraphics[scale=0.9]{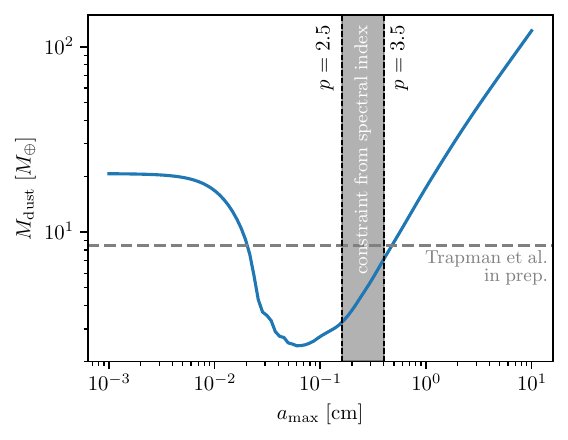}
    \caption{Estimated total dust mass for different disk-averaged maximum grain size models. The horizontal dashed line shows the estimated dust mass in Trapman et al. in prep. The vertical shaded area represents the $a_{\rm max}$ constraint estimated from the spectral index for a particle size distribution ranging from $p=$ 2.5 to 3.5.}
    \label{fig:DustMass}
\end{figure}

Constraining the average disk grain size is important because it helps in estimating the dust opacity and thus, the dust mass. Figure \ref{fig:DustMass} shows the total dust mass for different maximum sizes models. The total dust mass was computed using the de-convolved radial profile shown in Figure \ref{fig:continuum-analysis}, the DSHARP opacities for spherical compact grains, the scattering radiative transfer solution in \cite{Sierra_2020}, and assuming a dust temperature radial profile given by $T_d = [\zeta L_*/(4 \pi \sigma_B r^2)]^{1/4} $, where $\zeta = 0.02$ is the flaring angle \citep[e.g.,][]{Huang_2018}, $\sigma_B$ is the Stefan–Boltzmann constant, and $L_* = 0.25 L_{\sun}$ is the stellar luminosity \citep{Fang_2023}. The vertical shaded area shows the constraints on $a_{\rm max}$ from the spectral index between Band 7 and 6.

In addition, the horizontal dashed line is the total dust mass (M$_{\rm dust} = 8.5$ M$_{\oplus}$) estimated in Trapman et al. in prep., where optically thin emission, a dust opacity of 2.3 cm$^{2}$ g$^{-1}$, and an average disk temperature of $20$ K are assumed. This dust mass is consistent with averaged grain sizes of some hundred micrometers or some millimeter. However, the spectral index constraints support the estimation of millimeter grains.
Future dust continuum and molecular line observations at alternative millimeter wavelengths will be able to test the dust trap signatures around the J16120 ring.

\subsection{Compact dust emission, spiral arm structure, and bridge. Evidence of planet formation?} \label{sec:CPD}
We tentatively detected a compact source in dust continuum emission located inside the dust continuum gap. The source is detected at $\gtrsim 3\sigma$ level at peak. The peak flux within one beam centered at the compact source is $\sim 63 \ \mu$Jy. Assuming optically thin emission, a particle size distribution $n(a)da \propto a^{-3.5}da$, a temperature of 21 K (estimated from the stellar irradiation at the compact emission radius), and opacities $\kappa_{\nu} =$ 0.42 and 1.9 cm$^{2}$ g$^{-1}$ for 100 $\mu$m and 1 mm grain sizes \citep{Birnstiel_2018}, this flux corresponds to a dust mass of 12.9 and 2.7 Lunar masses for grain sizes of 100 $\mu$m and 1 mm, respectively. In PDS 70c, the estimated dust mass is between 2.6 and 0.6 Lunar masses for the same assumptions \citep{Benisty_2021}. In AS 209, the non-detection of millimeter emission at the location of the CPD candidate (200 au) sets an upper limit on the dust mass at $2.2$ Lunar masses, contrasting with the estimated CPD gas mass of $\sim 30 \ M_{\oplus}$ \citep{Bae_2022}. 

We propose that the compact source detected in the gap of J16120 is a CPD candidate. However, as pointed out by \cite{Andrews_2021}, it is imporantant to take into account that the distribution of pixel intensities within the gap tend to be very correlated because there are not enough independent resolution elements sampling this area. This artificially increase the rms in this region. In the J16120 gap, the rms is $\rm rms_{\rm gap}$ = 35 $\mu$Jy beam$^{-1}$ (a factor of 1.6 higher than the rms in and large empty area), which reduces the effective SNR of the compact source detection and makes it difficult to differentiate between random noise or genuine emission from a CPD.

In Appendix \ref{app:injection}, following the procedure in \cite{Andrews_2021},  we study the ability to detect a faint compact source in the gap of J16120, which is particularly important for high-contrast environments. The results show that the recovery fraction of a CPD with a flux similar to that of the compact emission is $\sim 20\%$. Deeper dust continuum observations of J16120 are needed to confirm or rule-out the possibility of the compact emission as a robust continuum detection.

On the other hand, we can also speculate that the tentative detection of a bridge connecting the ring with the inner disk (Figure \ref{fig:continuum-residual}) is a gap-crossing stream, constantly feeding the inner disk and fueling ongoing accretion into the star, as expected for a gap produced by a planetary mass object \citep[e.g.,][]{Price_2018}. Note that we do have some evidence of a gas gap (which spatially coincides with that from the dust continuum emission) for the optically thinner line emissions in Figure \ref{fig:observations}.

Although we show in Appendix \ref{app:Geom-Residuals} that the spiral residuals can be strongly affected by the uncertainties of the offset of the disk with respect to the phase center, our offset estimations minimize the residual structures and match with the brightest emission structures observed in the model-independent clean image. The extrapolation of the inferred spiral arm structure matches with the position of the compact dust emission and the bridge connecting the inner disk with the ring (Figure \ref{fig:continuum-residual}). Thus, it is not clear if these residuals are also part of the spiral arm structure or they are decoupled. Another possibility is that the spiral arm may be launched by a planet located at the compact emission, as it has been observed in numerical simulations and analytical studies \citep[e.g.,][]{Bae_2018}. To date, there is not additional observational evidence where a spiral arm structure is connected to a compact emission or CPD in a disk gap.

It is also important to remark that the emission from the positive residuals in the ring may originate from upper layers of the disk, where the observed emission is sensitive not only to dust surface density but to dust temperature structures. For example, \cite{Kiyoaki_2021} showed that the intensity of dust rings depends on the dust scale height for inclined disks. This geometrical effect may mimic azimuthal intensity asymmetries in the dust continuum maps. Our interpretation is only valid if the emission mainly originates from the disk midplane.
Meanwhile, the emission from the compact source and the bridge is presumably tracing the mid-plane, due to the deficit of mass in the gap. Future dust continuum observations tracing optically thicker and thinner wavelengths will allow us to disentangle the origin of all these disk substructures.

\subsection{A $^{12}$CO kink. Evidence of a circumplanetary disk candidate?} \label{sec:Discussion_CO}

The kinematic signature observed in the $^{12}$CO channel maps  of Figure \ref{fig:CO-channels} is similar to the kinematic perturbation expected from embedded planets in disks \citep[e.g.,][]{Perez_2015, Izquierdo_2021}, although disk stress profiles can also affect the velocity disk structure \citep{Rabago_2021}. Particularly, the channel at $4.3 \ \rm{km \ s}^{-1}$ shows a compact source close to the disk minor axis, beyond the dust continuum emission. This compact emission is connected to the inner region of the disk via an arm-like structure in the channel map at 4.2 $\rm{km \ s}^{-1}$. There are no similar kinematic structures below 4.0 $\rm{km \ s}^{-1}$ or above 4.3 $\rm{km \ s}^{-1}$, as shown in Figure \ref{fig:CO-extrachannels}.

An interesting velocity residual structure (within the green circle in the right panel of Figure \ref{fig:CO-modeling}) is close to the kinematic perturbations observed in the channel maps. However, there are residuals of the same order of magnitude outside this region, although none of them show a similar blue-white-red structure with approximately the same disk geometry. No similar velocity residuals are found when the velocity map is computed from gaussian fits to the line spectrum (Appendix \ref{app:velocity}), although they are not a good representation of line morphology close to the kink region. We can only speculate that this feature is tracing a rotational pattern at this position.

The center of this structure is localized at a deprojected radial position of $\sim 875$ mas (115.6 au), and at an azimuthal position of $\sim$ 130 deg. The size of the convolved structure has a diameter of $\diameter_{\rm extent} \sim 0.38^{\prime \prime}$ (50 au), with peak residuals of $v_{\rm peak} \approx \pm 0.17 \rm \ km \ s^{-1}$. 
Note that this is not the actual size, but an upper limit, as it is only marginally resolved. Even when de-convolved, this diameter is larger than typical values expected for circumplanetary disks.
However, if this were a rotating circumplanetary disk, the upper limit of the mass of the planet candidate (estimated from a Keplerian rotation) would be approximately $M_{\rm planet} \sim v_{\rm peak}^2 r_{\rm extent} /G \approx 0.8 \ M_{\rm Jupiter}$. The expected velocity deviations in the azimuthal profile for such a planet range between 2-3\% \citep{Armitage_2020}, which cannot be spectrally resolved given the spectral resolution of our observations (Table \ref{tab:Imaging}).

Planets candidates have been found via gas kinematics perturbations in several disks. The mass of the planet candidates is approximately $\sim 1 \ M_{\rm Jupiter}$ in HD 163296 \citep{Teague_2018, Izquierdo_2021} and $\sim 2 \ M_{\rm Jupiter}$ in HD 97048 \citep{Pinte_2019}. Additionally, masses of the order of a Jupiter mass have been estimated for planet candidates in DoAr 25, Elias 2-27, GW Lup, HD 143006, HD 163296, IM Lup, Sz 129, and WaOph 6 \citep{Pinte_2020}. All these planet mass estimations, derived from kinematic kinks, are higher than our upper limit estimate for J16120.
Less massive planets ($\lesssim 0.3 M_{\rm Jupiter}$), or those located near the disk minor axis, may not be detectable through kinematic perturbations, as they lack sufficient mass to significantly influence the gas dynamics, or may be camouflaged by the background velocities \citep{Izquierdo_2021}. In addition, not all kinks are necessarily associated with planets but with a Doppler flip, as in the disk around HD 100546 \citep{Casassus_2022}.
Higher resolution and sensitivity observations could help to support or discard our speculations about the planet candidate in the observed kink of J16120.

\section{Conclusions}\label{sec:conclusions}

As part of the ALMA Large Program ``AGE-PRO: ALMA survey of Gas Evolution in Protoplanetary disks", we present ALMA Band 6 observations of the disk around J16120. We image and model the dust continuum visibilities, and four detected molecular lines: $^{12}$CO (J=2-1), $^{13}$CO (J=2-1), C$^{18}$O (J=2-1), and H$_2$CO (J=$3_{(0,3)}-2_{(0,2)}$).
Both the continuum emission and molecular lines show a large gap and ring structure, with a peak at $\sim 0.57^{\prime \prime}$ (75.3 au). We find hints of planet formation signatures in the dust continuum and $^{12}$CO emission, including tentative detection of a compact dust continuum source in the disk gap, a dust continuum spiral arm, a bridge connection the outer ring with the inner disk, and a kink-like perturbation in the disk kinematics located outside of the dust ring. We summarize our findings below:
\begin{itemize}
    \item The resolved dust continuum observations of the J16120 disk reveal a wide ring, a large cavity, and a faint inner disk.
    We also detect a compact dust continuum source inside the large gap of the disk, at a deprojected radius of $0.24^{\prime \prime}$ (32 au). This compact emission is detected at a 3$\sigma$ level, with an estimated total dust mass between 12.9 and 2.7 Lunar masses (for grain sizes between 100$\mu$m and 1 mm, respectively). Although we propose this compact emission as a CPD candidate, the enhanced rms within the gap makes it difficult to distinguish this emission from local thermal noise.

    \item The non-axisymmetric structures in the dust continuum emission of J16120 show an excess emission localized at 1) the CPD candidate, 2) a bridge  connecting the inner disk and the ring, and 3) within the South-Western part of the ring.
    The residuals within the ring follows a spiral arm structure, and its inward extrapolation matches with the spatial position of the CPD candidate and the bridge.

    \item Moment 0 maps of the detected molecular lines show a ring like structure, localized around the radial position of the dust continuum peak. The $^{13}$CO, C$^{18}$O, and H$_2$CO radial profiles show an emission cavity within the ring, while the $^{12}$CO emission shows the presence of gas within the cavity.

    \item Peak moment maps of the detected molecular lines (including $^{12}$CO) show a clear ring like structure, with a peak around the radial position of the dust continuum ring. These maps suggest a gas gap that spatially coincides with that observed in the dust continuum emission. In addition, a bright emission structure is observed in the North East of all the CO molecular lines, which could be tracing high temperatures of gas surface densities.
    
    \item The $^{12}$CO channel maps between 4.0 and 4.3 $\mathrm{km \ s}^{-1}$ show a kinematic perturbation towards the South of the disk, at a deprojected radius of $0.85^{\prime \prime}$ (112 au). We did not detect similar kinematic perturbations in the CO isotopologues and H$_2$CO channel maps, presumable due to their limited angular resolution and sensitivity.

    \item The modeling of the $^{12}$CO velocity map around J16120 helps us to constrain a stellar mass of 0.7 $M_{\odot}$, an emission scale height of 0.17$^{\prime \prime}$ (22.5 au) at a radius of 1$^{\prime \prime}$ (132 au), and a taper radius of $1.3^{\prime \prime}$ (172 au).

    \item After subtracting a best velocity model to the line of sight velocity map, we observed a rotating-like structure close to the kink location, with a position angle and inclination similar to that of the disk. Similar order of magnitude residuals are found beyond the kink area, but none of them show a gradient with approximately the same disk geometry. We speculate that these residuals could be tracing a rotating structure close to the kink location.

    \item The $^{12}$CO kinematics reveals an azimuthal velocity ($v_{\phi}$) that matches (within the error bars) with the expected Keplerian velocity in the midplane. Thus, it is not clear if the origin of the dust continuum ring is a dust trap created by a pressure bump. 
\end{itemize}

Future infrared and millimeter observations (e.g., ALMA Band 7, Project Code 2023.1.01100.S in progress) will help in further understanding, confirmation, and identification of the planet formation signatures in the disk around J16120. Observational evidence is crucial for determining when planet formation around low-mass stars begins and understanding the physical properties of forming planets.

\acknowledgments
This paper makes use of the following ALMA data: ADS/JAO.ALMA\#2021.1.00128.L ALMA is a partnership of ESO (representing its member states), NSF (USA) and NINS (Japan), together with NRC (Canada), MOST and ASIAA (Taiwan), and KASI (Republic of Korea), in cooperation with the Republic of Chile. The Joint ALMA Observatory is operated by ESO, AUI/NRAO and NAOJ. The National Radio Astronomy Observatory is a facility of the National Science Foundation operated under cooperative agreement by Associated Universities, Inc.
Acknowledges from each AGE-PRO member will be included in the following versions.

A.S.\ acknowledges support from FONDECYT Postdoctorado 2022 \#3220495.
A.S.\ and P.P.\ acknowledge funding from the UK Research and Innovation (UKRI) under the UK government’s Horizon Europe funding guarantee from ERC (under grant agreement No 101076489).
L.P.\ acknowledges support from ANID BASAL project FB210003 and ANID FONDECYT Regular grant \#1221442.
L.P.\ and C.A.G.\ acknowledge support from ESO-Chile Comité Mixto 2023. 
C.A.G.\ acknowledges support from FONDECYT Postdoctorado 2021 \#3210520. 
J.M. acknowledges support from FONDECYT de Postdoctorado 2024 \#3240612 and from the Millennium Nucleus on Young Exoplanets and their Moons (YEMS), ANID - Center Code  NCN2021\_080.
K.Z.\ acknowledges the support of the NSF AAG grant \#2205617.
L.T.\ acknowledges the support of NSF AAG grant \#2205617.
L.A.C.\ acknowledges support from FONDECYT grant \#1241056.
C.G-R. and L.A.C.\ acknowledge support from the Millennium Nucleus on Young Exoplanets and their Moons (YEMS), ANID - Center Code NCN2021\_080.
G.R.\ acknowledges funding from the Fondazione Cariplo, grant no. 2022-1217, and the European Research Council (ERC) under the European Union’s Horizon Europe Research \& Innovation Programme under grant agreement no. 101039651 (DiscEvol). Views and opinions expressed are however those of the author(s) only, and do not necessarily reflect those of the European Union or the European Research Council Executive Agency. Neither the European Union nor the granting authority can be held responsible for them.
K.S.\ acknowledges support from the European Research Council under the Horizon 2020 Framework Program via the ERC Advanced Grant Origins 83 24 28.
B.T.\ acknowledges the support of  the Programme National Physique et Chimie du Milieu Interstellaire (PCMI) of CNRS/INSU with INC/INP co-funded by CEA and CNES. 
I.P. and D.D. acknowledge support from Collaborative NSF Astronomy \& Astrophysics Research grant (ID: 2205870)

\software{Astropy \citep{astropy:2013, astropy:2018}, bettermoments \citep{Teague_2018b}, CASA \citep{McMullin_2007}, Corner \citep{corner}, Eddy \citep{eddy}, Frankenstein \citep{Jennings_2020}, Matplotlib \citep{Matplotlib_2007}, Numpy \citep{Numpy_2020}}

\newpage
\appendix
\section{Supporting figures} \label{app:Supporting-Figs}
\begin{figure}
    \centering
    \includegraphics[width=\textwidth]{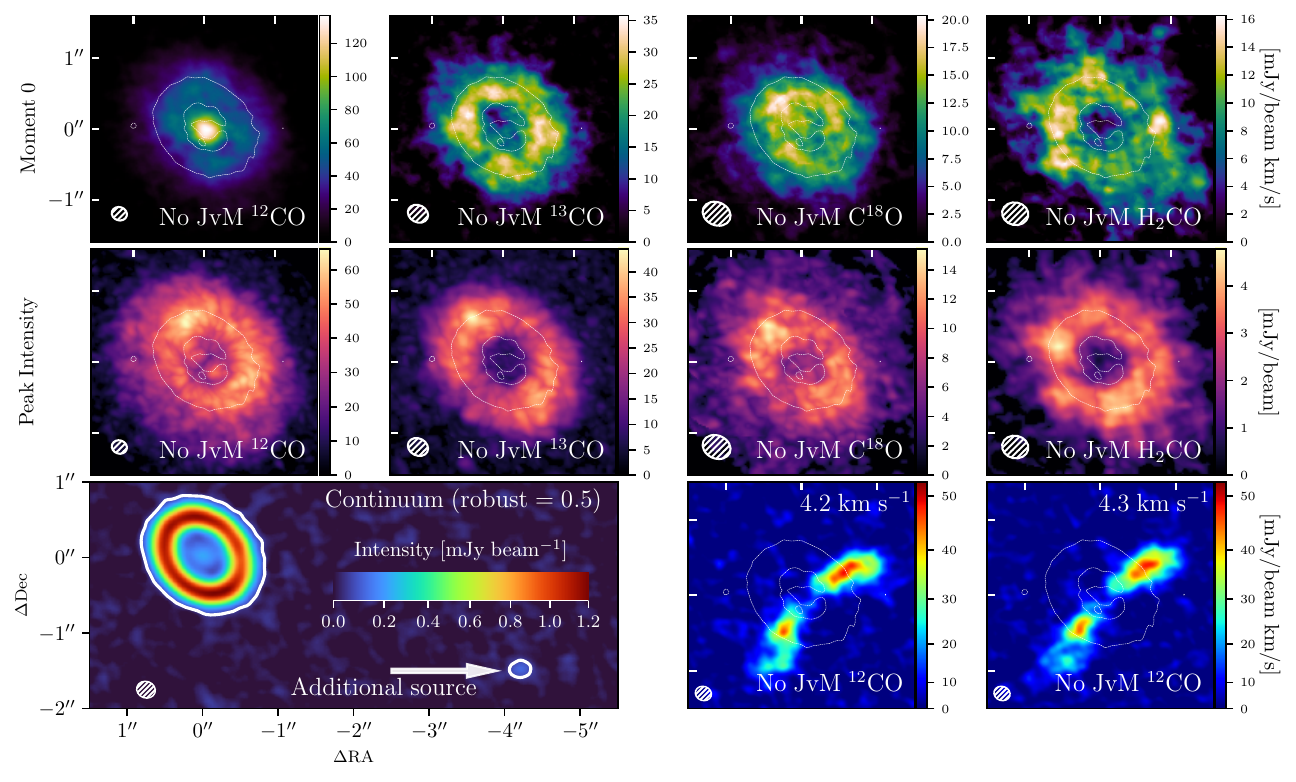}
    \caption{Top panels from left to right: $\rm ^{12}CO \ (J=2-1)$, $\rm ^{13}CO \ (J=2-1)$, $\rm C^{18}O \ (J=2-1)$, and $\rm H_2CO \ (J=3_{(0,3)}-2_{(0,2)})$ moment 0 maps without JvM correction.
    Middle panels: Peak intensity maps for the same molecular lines without JvM correction. Bottom left panel: Dust continuum emission map with a Briggs robust parameter = 0.5 and an additional faint source detection. Contours levels are 4$\sigma$ in this panel.  Bottom right panels: $^{12}$CO emission channel maps at 4.2 and 4.3 km s$^{-1}$ without JvM correction. Dust continuum iso-contours at 3$\sigma$ from Figure \ref{fig:continuum-analysis} are shown in all panels with molecular line data.}
    \label{fig:Supporting-Figures}
\end{figure}

\begin{figure}
    \centering
    \includegraphics[scale=1.0]{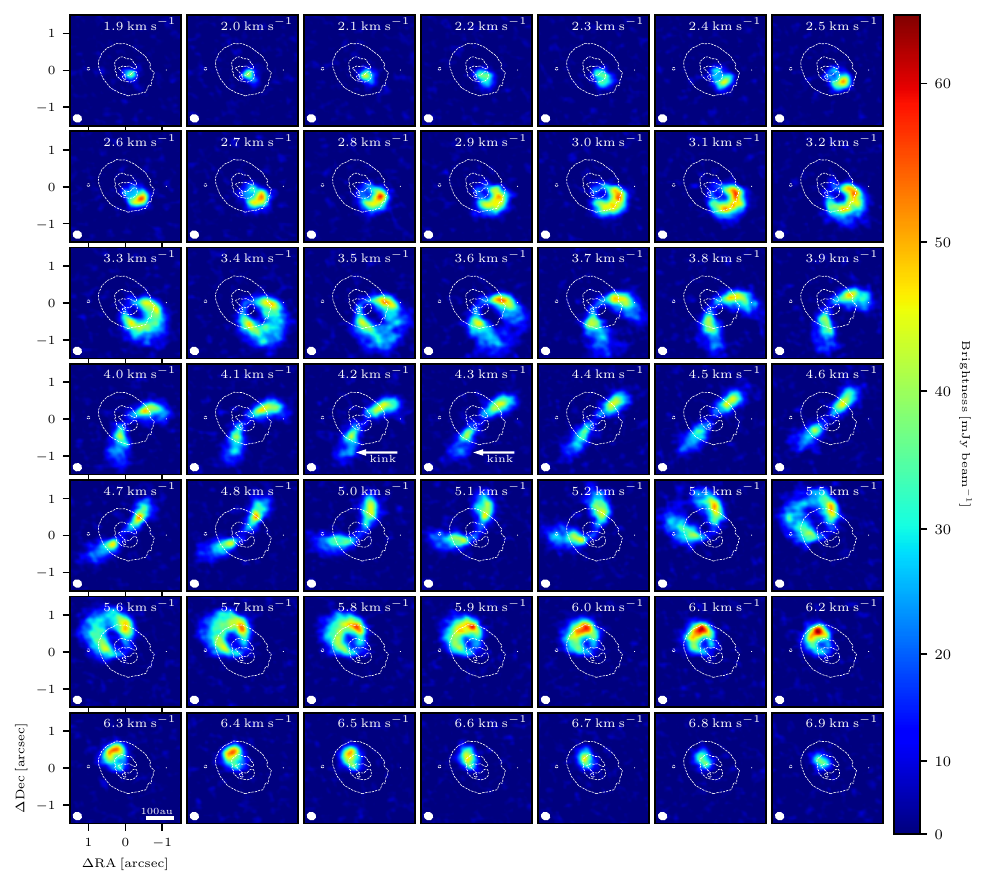}
    \caption{Additional channel maps of the $^{12}$CO emission around J16120. The color bar in the right is the same for all panels. The synthesized beam is shown in the bottom left panels, and a horizontal line with a size of 100 au is shown in the last row and first column panel.}
    \label{fig:CO-extrachannels}
\end{figure}

Figure \ref{fig:Supporting-Figures} shows the moment 0 maps and two selective channel maps where JvM correction was not applied. The ring structure, and velocity deviations in these maps are similar to those shown in Figures \ref{fig:observations} and \ref{fig:CO-channels}, showing that the JvM correction is not creating artificial disk structures. Additionally, the bottom left panel shows a larger map of the dust continuum emission, where a faint additional source is detected. Figure \ref{fig:CO-extrachannels} shows additional $^{12}$CO emission channel maps between 1.9 and 6.9 km s$^{-1}$. The full cube was imaged between -5.5 and 14.4 km s$^{-1}$.

\section{More on dust continuum modeling}
\subsection{Effects of geometry on the residual maps} \label{app:Geom-Residuals}
\begin{figure}
    \centering
    \includegraphics[width=\textwidth]{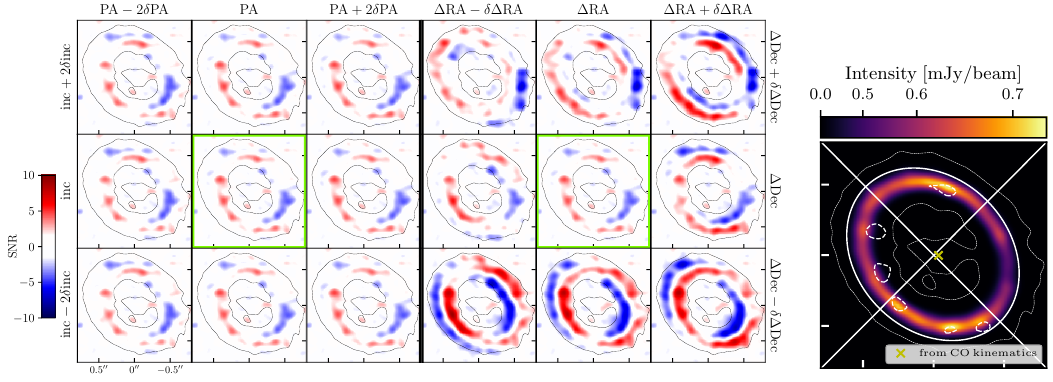}
    \caption{\textbf{Left:} Effects of geometry and offset on the residual continuum maps. The first three columns show the effects of the inclination (inc) and position angle (PA), and the last three columns show the effect of the disk offset with respect to the phase center ($\mathrm{\Delta RA, \Delta Dec}$). The central panel highlighted in green is the residual map corresponding to the central values of geometry and offset presented in Figure \ref{fig:continuum-residual}. The iso-contour in all panels is taken from Figure \ref{fig:continuum-analysis} at 3$\sigma$ level. \textbf{Right:} Dust continuum clean image. The color bar was adjusted to visualize the high intensity emission. The solid lines show the disk minor and major axis, while the solid ellipse show an ellipse computed from the position angle and disk inclination and centred at the disk offset. The yellow cross shows the disk center computed from the $^{12}$CO kinematics.    
    The white dashed lines are the 3$\sigma$ iso-contours from the residual map of the central panels in green of the left figure.}
    \label{fig:geom-errors}
\end{figure}

Errors on the disk inclination (inc), position angle (PA), or the offset of the disk center with respect to the phase center ($\mathrm{\Delta RA, \Delta Dec}$) can lead to artificial residual features when an axi-symmetric disk model is subtracted from the data \citep{Andrews_2021}.
We study the effects of these parameters on the dust continuum residual maps by varying the disk geometry and offset within the error bars computed in Section \ref{sec:continuum}. The left side of Figure \ref{fig:geom-errors} shows the residual maps for different inclination and position angles (the first three columns in the left), and the disk center offset with respect to the phase center (the last three columns in the right). 

The effects of varying the inclination and position angle within twice the error bars are almost indistinguishable from the best-fit values (central panel in green). However, the effects of the offset on the residual maps are more significant. Symmetric positive and negative residual structures appear when the disk center is spatially shifted in any direction. These artificial structures merge with the spiral residuals inferred from the best fit values (central panel in green). The error bars in right ascension and declination are 20, 10 mas respectively, corresponding to a pixel size and half a pixel size of the 20 mas cellsize of the clean image. Note that although the spiral arms residuals can be strongly affected by the disk offset, the residuals of the compact emission and the bridge are only slightly different, confirming a robust detection of these structures. 

Although offset effects can have important effects on the residual maps, our best fit for the geometry and offset minimize the residual structures. A more sophisticated disk morphology model could be able  to explain the residuals observed in the dust continuum map, as the eccentricity model of the circumbinary disk around CS Cha \citep{Kurtovic_2022}.
The residual maps in this work only show the deviations from axi-symmetric structures. These asymmetries can be observed directly in the clean image data, independently of the constraints on geometry and offset and the subtracted model, as shown in the right side of Figure \ref{fig:geom-errors}. In this image, the color bar has been adjusted to better visualize the high-intensity morphologies of the disk. The spatial location of the dust continuum residuals coincides with the positions of bright structures observed in the dust continuum map, confirming the presence of these non axi-symmetric structures. The asymmetries are detected at $3\sigma$ level $\sim 0.06$ mJy/beam in a ring with an intensity of $\sim 0.6$ mJy/beam, resulting in a contrast of $\sim 10\%$ in the dust continuum map.
The right figure also shows a yellow cross with the disk center coordinates computed from the CO kinematic model fit in Section \ref{sec:lineCO}. This center is West of the dust continuum center, with a separation of $\sim 40$ mas (2 pixels or $\sim 1/4$ the dust continuum beam size). The difference in declination is negligible.

\subsection{CPD injection tests}\label{app:injection}
Imaging and detecting faint sources in high-contrast environments is challenging, requiring the assessment of our ability to detect them in the dust continuum data of this work. To address this study, we first remove the emission of the compact structure and the bridge presented in Figure \ref{fig:continuum-analysis}. This is done by isolating and computing the visibilities of the model image (clean components) around the compact emission and the bridge region, and then subtracting them from the original data visibilities. The visibilities of the compact emission and the bridge are computed with \textsc{Galario} \citep {Tazzari_2018} at the same observed spatial spatial frequencies $(u,v)$.

\begin{figure}
    \centering
    \includegraphics[width=0.5\textwidth]{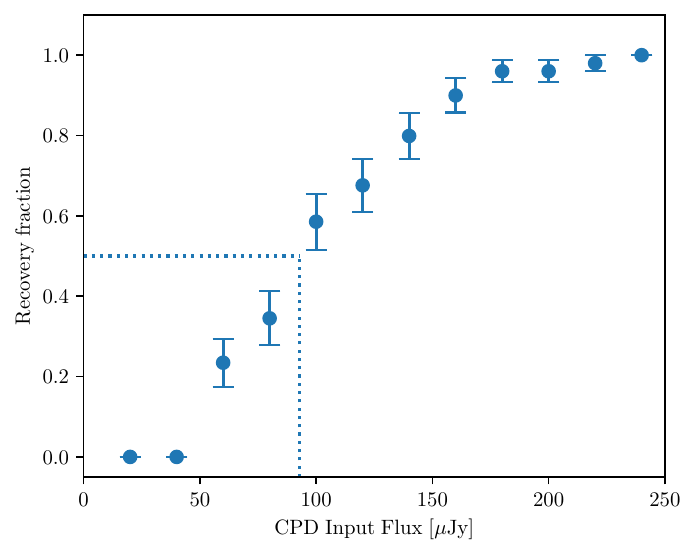}
    \caption{Fraction of CPD fluxes that are recovered as a function of the CPD input flux within the gap of J16120. 
    The 50\% recovery fraction (horizontal dotted line) corresponds to $93\mu$Jy (vertical dotted line).}
    \label{fig:Injection}
\end{figure}

Once the gap is free of compact sources, we follow the procedure described in \cite{Andrews_2021}. We created 50 different data set models where we injected the visibilities of a CPD with different fluxes at random coordinates within the gap. Then, the visibilities from a Frank model are subtracted from the data and the residuals visibilities are imaged with \textsc{CASA}. The peak emission within the gap region is measured and compared with the input flux model. If the peak emission is equal or higher than the input flux, the detection is a ``success''.
Figure \ref{fig:Injection} shows the recovery fraction as a function of the CPD input flux model. For a CPD with a flux similar to the flux of the compact emission ($\sim 63 \mu$Jy), the recovery fraction is $\sim 0.2$. The interpolated 50\% fraction corresponds to 93 $\mu$Jy.

\section{More on kinematics modeling}
\subsection{Isotopologue dynamics}
\label{app:MoreKinematics}
\begin{figure}
    \centering
    \includegraphics[scale=0.35]{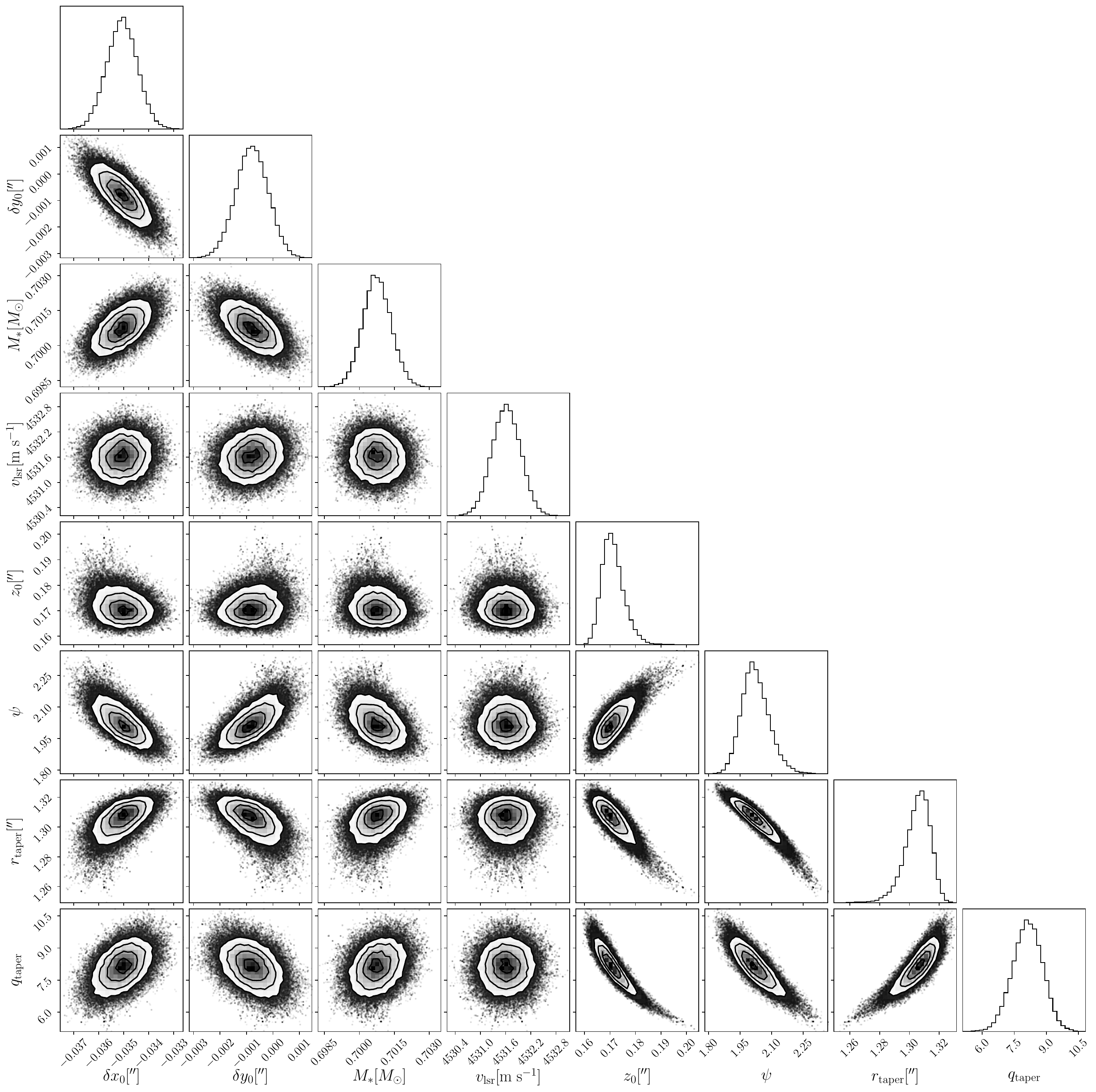}
    \caption{Corner plots of the last 3000 steps of the Markov Monte Carlo chain used to explore the $^{12}$CO emission surface parameters.}
    \label{fig:CornerPlotCO}
\end{figure}

Figure \ref{fig:CornerPlotCO} shows the corner plots of the emission surface fit using the $^{12}$CO kinematics and \textsc{eddy} (Section \ref{sec:lineCO}). The corner plots were computed using the last 3000 steps of the Markov Monte Carlo chain, and the 64 walkers used to explore the parameters of the $^{12}$CO emission surface (Equation \ref{eq:EmissionSurface}).
The space parameter is initially explored using 3000 steps for each walker, but they are finally discarded to create the corner plots.

\begin{figure}
    \centering
    \includegraphics[width=0.8\textwidth]{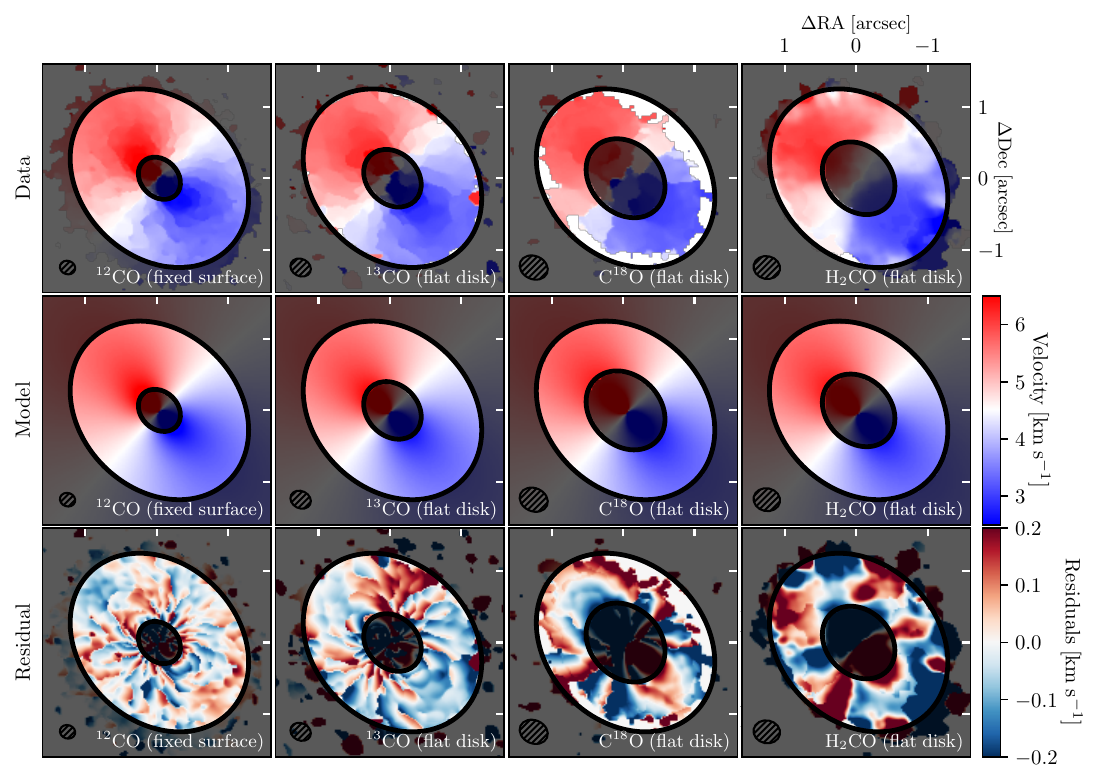}
    \caption{Kinematic modeling of CO (first column), $^{13}$CO (second column), C$^{18}$O (third column) and H$_2$CO (fourth column). The first row shows the velocity field maps, the second row is the best fit velocity model, and the third row are the velocity residuals.}
    \label{fig:Isotopologues}
\end{figure}

Figure \ref{fig:Isotopologues} shows the kinematic modeling for the three detected CO isotopologues and H$_2$CO. For $^{12}$CO, the emission surface is given by the parametric fit obtained from emission surface measurements (yellow line in Figure \ref{fig:CO-surface}, respectively), in contrast to the emission surface parametrization (red line in Figure \ref{fig:CO-surface}) obtained from the Keplerian fit in Figure \ref{fig:CO-modeling}. As discussed in Section \ref{sec:lineCO}, there are not important differences between both $^{12}$CO emission surface models. The residual structure highlighted with a green circle in Figure \ref{fig:CO-modeling} is also shown in the top right panel in Figure \ref{fig:Isotopologues}.

The kinematics models for $^{13}$CO, C$^{18}$O, and H$_2$CO in Figure \ref{fig:Isotopologues} are flat disks (no emission surface is fitted), as discussed in Section \ref{sec:line_isotop}. The best fit parameters for each keplerian model are summarized in Table \ref{tab:Surface}.

\subsection{Line of sight velocity maps and modeling} \label{app:velocity}
\begin{figure}
    \centering
    \includegraphics[width=\textwidth]{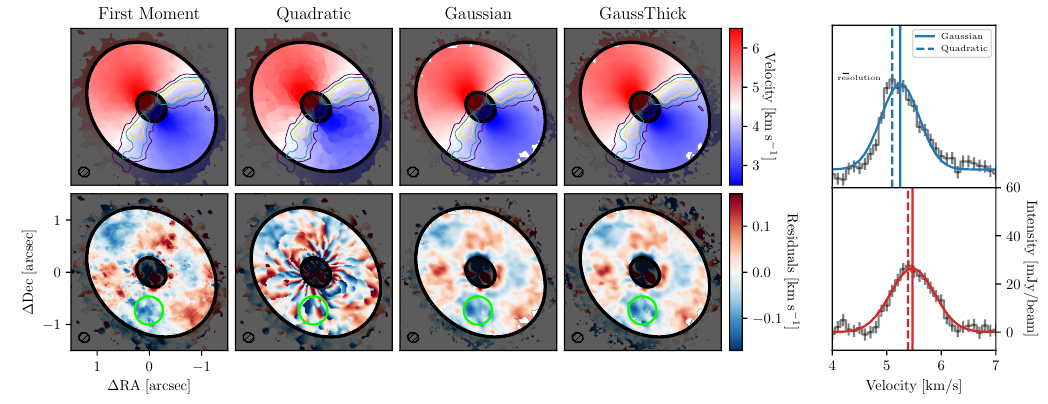} 
    \caption{Maps in the top left panels: Different estimation of the line of sight velocity. From left to right: \texttt{First Moment, Quadratic, Gaussian, GaussThick}. Maps in the bottom left panels: Residuals after subtracting the best fit model of each velocity map. Right panels: Molecular line spectrum at two pixels within the green circle of the left panels. The vertical lines are the line of sight velocities estimated from the gaussian (solid line) and quadratic (dashed line) methodologies. The spectral resolution is shown as a horizontal line in the top panel.}
    \label{fig:velocities}
\end{figure}

The line-of-sight velocity has traditionally been estimated using the moment 1 map (intensity-weighted average velocity). However, it is known that the moment 1 map is strongly affected by noise in the spectrum and any asymmetry in the line profile \citep{Teague_2018b}. Here we show velocity maps computed by alternative methodologies: \texttt{Quadratic, Gaussian, GuassThick}, which are part of the \textsc{bettermoments} tools described in \cite{Teague_2018b}. 
The maps shown in the top row of Figure \ref{fig:velocities} show the line of sight velocities computed from different estimations. The maps in the bottom panels show the residuals when a best fit model is subtracted for each velocity map. The best fit model for each velocity map is computed independently. Table \ref{tab:Velocities} shows the best fit parameters for each case.

The differences between the line of sight velocities computed from the \texttt{Quadratic} and \texttt{Gaussian-GaussThick} methodologies are smaller than the spectral resolution ($0.1$ km s$^{-1}$) in most of the disk. However, they differ by $\gtrsim 0.15$ km s$^{-1}$ in the region close to the kink location, where the line spectrum looks asymmetric and the gaussian fit is not a good representation of the line emission, as shown in the right panels for two line spectrum within the green circle of the left panels. The \texttt{Quadratic} method has been used in several works \citep[e.g.,][]{Yu_2021, Galloway_2023, Zhao_2024}, due to it has been proved to be a robust methodology to estimate the line of sight velocity.
Note the residual structure observed within the green circle of the \texttt{Quadratic} method does not appear in the other cases where the line of sight is not well estimated.

\begin{table*}
    \centering
    \caption{Bet fit parameters for $^{12}$CO velocity maps computed by different methodologies.}    
    \begin{tabular}{c|cccccccc}
    \hline \hline
    Method                & $\delta x_0$ &  $\delta y0$ & $M_*$ & $v_{\rm lsr}$ & $z_0$ & $\psi$ & $r_{\rm taper}$ & $q_{\rm taper}$\\ 
     & [mas] & [mas] & [$M_{\odot}$] & [km s$^{-1}$] & [arcsec] & & [arcsec] & \\
    \hline
    \texttt{First Moment} & $10^{+10}_{-40}$ & $-54^{+55}_{-11}$ & $0.76^{+0.3}_{-0.4}$ & $4.5^{+0.1}_{-0.1}$ & $0.17^{*}$ & $1.0^{+5}_{-2}$ & $0.2^{+5.}_{-0.2}$ & $11.1^{+20.3}_{-10.3}$  \\
    \texttt{Quadratic}    & $-35.0^{+1.0}_{-1.1}$ & $-1.0^{+1.0}_{-1.0}$ & $0.70 ^{+0.1}_{-0.1}$ & $4.5^{+0.1}_{-0.1}$ & $0.17^{+0.01}_{-0.01}$ & $2.0^{+0.1}_{-0.1}$ & $1.3^{+0.2}_{-0.2}$ & $8.3^{+0.8}_{-0.7}$ \\
    \texttt{Gaussian}     & $-36.6^{+0.6}_{-1.0}$ & $-8.0^{+0.8}_{-1.0}$ & $0.70^{+0.1}_{-0.1}$ & $4.5^{+0.2}_{-0.1}$ & $0.17^{*}$ & $2.6^{+0.1}_{-0.1}$ & $1.1^{+0.1}_{-0.1}$ & $5.4^{+0.2}_{-0.3}$ \\
    \texttt{GaussThick}   & $-36.5^{+0.8}_{-0.9}$ & $-7.7^{+0.7}_{-0.8}$ & $0.70^{+0.1}_{-0.1}$ & $4.5^{+0.1}_{-0.1}$ & $0.17^{*}$ & $2.6^{+0.2}_{-0.1}$ & $1.1^{+0.1}_{-0.1}$ & $5.4^{+0.2}_{-0.2}$\\
    \hline \hline
    \end{tabular}
    $^{*}$Fixed parameters during the fitting.
    \label{tab:Velocities}
\end{table*}

\bibliography{main}{}

\begin{thebibliography}{}
\expandafter\ifx\csname natexlab\endcsname\relax\def\natexlab#1{#1}\fi
\providecommand{\url}[1]{\href{#1}{#1}}
\providecommand{\dodoi}[1]{doi:~\href{http://doi.org/#1}{\nolinkurl{#1}}}
\providecommand{\doeprint}[1]{\href{http://ascl.net/#1}{\nolinkurl{http://ascl.net/#1}}}
\providecommand{\doarXiv}[1]{\href{https://arxiv.org/abs/#1}{\nolinkurl{https://arxiv.org/abs/#1}}}

\bibitem[{{Alexander} {et~al.}(2014){Alexander}, {Pascucci}, {Andrews},
  {Armitage}, \& {Cieza}}]{Alexander_2014}
{Alexander}, R., {Pascucci}, I., {Andrews}, S., {Armitage}, P., \& {Cieza}, L.
  2014, in Protostars and Planets VI, ed. H.~{Beuther}, R.~S. {Klessen}, C.~P.
  {Dullemond}, \& T.~{Henning}, 475--496,
  \dodoi{10.2458/azu_uapress_9780816531240-ch021}

\bibitem[{{ALMA Partnership} {et~al.}(2015){ALMA Partnership}, {Brogan},
  {P{\'e}rez}, {Hunter}, {Dent}, {Hales}, {Hills}, {Corder}, {Fomalont},
  {Vlahakis}, {Asaki}, {Barkats}, {Hirota}, {Hodge}, {Impellizzeri}, {Kneissl},
  {Liuzzo}, {Lucas}, {Marcelino}, {Matsushita}, {Nakanishi}, {Phillips},
  {Richards}, {Toledo}, {Aladro}, {Broguiere}, {Cortes}, {Cortes}, {Espada},
  {Galarza}, {Garcia-Appadoo}, {Guzman-Ramirez}, {Humphreys}, {Jung}, {Kameno},
  {Laing}, {Leon}, {Marconi}, {Mignano}, {Nikolic}, {Nyman}, {Radiszcz},
  {Remijan}, {Rod{\'o}n}, {Sawada}, {Takahashi}, {Tilanus}, {Vila Vilaro},
  {Watson}, {Wiklind}, {Akiyama}, {Chapillon}, {de Gregorio-Monsalvo}, {Di
  Francesco}, {Gueth}, {Kawamura}, {Lee}, {Nguyen Luong}, {Mangum}, {Pietu},
  {Sanhueza}, {Saigo}, {Takakuwa}, {Ubach}, {van Kempen}, {Wootten},
  {Castro-Carrizo}, {Francke}, {Gallardo}, {Garcia}, {Gonzalez}, {Hill},
  {Kaminski}, {Kurono}, {Liu}, {Lopez}, {Morales}, {Plarre}, {Schieven},
  {Testi}, {Videla}, {Villard}, {Andreani}, {Hibbard}, \&
  {Tatematsu}}]{ALMA_2015}
{ALMA Partnership}, {Brogan}, C.~L., {P{\'e}rez}, L.~M., {et~al.} 2015, \apjl,
  808, L3, \dodoi{10.1088/2041-8205/808/1/L3}

\bibitem[{{Andrews}(2020)}]{Andrews_2020}
{Andrews}, S.~M. 2020, \araa, 58, 483,
  \dodoi{10.1146/annurev-astro-031220-010302}

\bibitem[{{Andrews} {et~al.}(2012){Andrews}, {Wilner}, {Hughes}, {Qi},
  {Rosenfeld}, {{\"O}berg}, {Birnstiel}, {Espaillat}, {Cieza}, {Williams},
  {Lin}, \& {Ho}}]{Andrews_2012}
{Andrews}, S.~M., {Wilner}, D.~J., {Hughes}, A.~M., {et~al.} 2012, \apj, 744,
  162, \dodoi{10.1088/0004-637X/744/2/162}

\bibitem[{{Andrews} {et~al.}(2018){Andrews}, {Huang}, {P{\'e}rez}, {Isella},
  {Dullemond}, {Kurtovic}, {Guzm{\'a}n}, {Carpenter}, {Wilner}, {Zhang}, {Zhu},
  {Birnstiel}, {Bai}, {Benisty}, {Hughes}, {{\"O}berg}, \&
  {Ricci}}]{Andrews_2018}
{Andrews}, S.~M., {Huang}, J., {P{\'e}rez}, L.~M., {et~al.} 2018, \apjl, 869,
  L41, \dodoi{10.3847/2041-8213/aaf741}

\bibitem[{{Andrews} {et~al.}(2021){Andrews}, {Elder}, {Zhang}, {Huang},
  {Benisty}, {Kurtovic}, {Wilner}, {Zhu}, {Carpenter}, {P{\'e}rez}, {Teague},
  {Isella}, \& {Ricci}}]{Andrews_2021}
{Andrews}, S.~M., {Elder}, W., {Zhang}, S., {et~al.} 2021, \apj, 916, 51,
  \dodoi{10.3847/1538-4357/ac00b9}

\bibitem[{{Ansdell} {et~al.}(2018){Ansdell}, {Williams}, {Trapman}, {van
  Terwisga}, {Facchini}, {Manara}, {van der Marel}, {Miotello}, {Tazzari},
  {Hogerheijde}, {Guidi}, {Testi}, \& {van Dishoeck}}]{Ansdell_2018}
{Ansdell}, M., {Williams}, J.~P., {Trapman}, L., {et~al.} 2018, \apj, 859, 21,
  \dodoi{10.3847/1538-4357/aab890}

\bibitem[{{Astropy Collaboration} {et~al.}(2013){Astropy Collaboration},
  {Robitaille}, {Tollerud}, {Greenfield}, {Droettboom}, {Bray}, {Aldcroft},
  {Davis}, {Ginsburg}, {Price-Whelan}, {Kerzendorf}, {Conley}, {Crighton},
  {Barbary}, {Muna}, {Ferguson}, {Grollier}, {Parikh}, {Nair}, {Unther},
  {Deil}, {Woillez}, {Conseil}, {Kramer}, {Turner}, {Singer}, {Fox}, {Weaver},
  {Zabalza}, {Edwards}, {Azalee Bostroem}, {Burke}, {Casey}, {Crawford},
  {Dencheva}, {Ely}, {Jenness}, {Labrie}, {Lim}, {Pierfederici}, {Pontzen},
  {Ptak}, {Refsdal}, {Servillat}, \& {Streicher}}]{astropy:2013}
{Astropy Collaboration}, {Robitaille}, T.~P., {Tollerud}, E.~J., {et~al.} 2013,
  \aap, 558, A33, \dodoi{10.1051/0004-6361/201322068}

\bibitem[{{Astropy Collaboration} {et~al.}(2018){Astropy Collaboration},
  {Price-Whelan}, {Sip{\H{o}}cz}, {G{\"u}nther}, {Lim}, {Crawford}, {Conseil},
  {Shupe}, {Craig}, {Dencheva}, {Ginsburg}, {Vand erPlas}, {Bradley},
  {P{\'e}rez-Su{\'a}rez}, {de Val-Borro}, {Aldcroft}, {Cruz}, {Robitaille},
  {Tollerud}, {Ardelean}, {Babej}, {Bach}, {Bachetti}, {Bakanov}, {Bamford},
  {Barentsen}, {Barmby}, {Baumbach}, {Berry}, {Biscani}, {Boquien}, {Bostroem},
  {Bouma}, {Brammer}, {Bray}, {Breytenbach}, {Buddelmeijer}, {Burke},
  {Calderone}, {Cano Rodr{\'\i}guez}, {Cara}, {Cardoso}, {Cheedella}, {Copin},
  {Corrales}, {Crichton}, {D'Avella}, {Deil}, {Depagne}, {Dietrich}, {Donath},
  {Droettboom}, {Earl}, {Erben}, {Fabbro}, {Ferreira}, {Finethy}, {Fox},
  {Garrison}, {Gibbons}, {Goldstein}, {Gommers}, {Greco}, {Greenfield},
  {Groener}, {Grollier}, {Hagen}, {Hirst}, {Homeier}, {Horton}, {Hosseinzadeh},
  {Hu}, {Hunkeler}, {Ivezi{\'c}}, {Jain}, {Jenness}, {Kanarek}, {Kendrew},
  {Kern}, {Kerzendorf}, {Khvalko}, {King}, {Kirkby}, {Kulkarni}, {Kumar},
  {Lee}, {Lenz}, {Littlefair}, {Ma}, {Macleod}, {Mastropietro}, {McCully},
  {Montagnac}, {Morris}, {Mueller}, {Mumford}, {Muna}, {Murphy}, {Nelson},
  {Nguyen}, {Ninan}, {N{\"o}the}, {Ogaz}, {Oh}, {Parejko}, {Parley}, {Pascual},
  {Patil}, {Patil}, {Plunkett}, {Prochaska}, {Rastogi}, {Reddy Janga},
  {Sabater}, {Sakurikar}, {Seifert}, {Sherbert}, {Sherwood-Taylor}, {Shih},
  {Sick}, {Silbiger}, {Singanamalla}, {Singer}, {Sladen}, {Sooley},
  {Sornarajah}, {Streicher}, {Teuben}, {Thomas}, {Tremblay}, {Turner},
  {Terr{\'o}n}, {van Kerkwijk}, {de la Vega}, {Watkins}, {Weaver}, {Whitmore},
  {Woillez}, {Zabalza}, \& {Astropy Contributors}}]{astropy:2018}
{Astropy Collaboration}, {Price-Whelan}, A.~M., {Sip{\H{o}}cz}, B.~M., {et~al.}
  2018, \aj, 156, 123, \dodoi{10.3847/1538-3881/aabc4f}

\bibitem[{{Bae} {et~al.}(2023){Bae}, {Isella}, {Zhu}, {Martin}, {Okuzumi}, \&
  {Suriano}}]{Bae_2023}
{Bae}, J., {Isella}, A., {Zhu}, Z., {et~al.} 2023, in Astronomical Society of
  the Pacific Conference Series, Vol. 534, Protostars and Planets VII, ed.
  S.~{Inutsuka}, Y.~{Aikawa}, T.~{Muto}, K.~{Tomida}, \& M.~{Tamura}, 423,
  \dodoi{10.48550/arXiv.2210.13314}

\bibitem[{{Bae} \& {Zhu}(2018)}]{Bae_2018}
{Bae}, J., \& {Zhu}, Z. 2018, \apj, 859, 118, \dodoi{10.3847/1538-4357/aabf8c}

\bibitem[{{Bae} {et~al.}(2022){Bae}, {Teague}, {Andrews}, {Benisty},
  {Facchini}, {Galloway-Sprietsma}, {Loomis}, {Aikawa}, {Alarc{\'o}n},
  {Bergin}, {Bergner}, {Booth}, {Cataldi}, {Cleeves}, {Czekala}, {Guzm{\'a}n},
  {Huang}, {Ilee}, {Kurtovic}, {Law}, {Le Gal}, {Liu}, {Long}, {M{\'e}nard},
  {{\"O}berg}, {P{\'e}rez}, {Qi}, {Schwarz}, {Sierra}, {Walsh}, {Wilner}, \&
  {Zhang}}]{Bae_2022}
{Bae}, J., {Teague}, R., {Andrews}, S.~M., {et~al.} 2022, \apjl, 934, L20,
  \dodoi{10.3847/2041-8213/ac7fa3}

\bibitem[{{Benisty} {et~al.}(2021){Benisty}, {Bae}, {Facchini}, {Keppler},
  {Teague}, {Isella}, {Kurtovic}, {P{\'e}rez}, {Sierra}, {Andrews},
  {Carpenter}, {Czekala}, {Dominik}, {Henning}, {Menard}, {Pinilla}, \&
  {Zurlo}}]{Benisty_2021}
{Benisty}, M., {Bae}, J., {Facchini}, S., {et~al.} 2021, \apjl, 916, L2,
  \dodoi{10.3847/2041-8213/ac0f83}

\bibitem[{{Birnstiel}(2023)}]{Birnstiel_2023}
{Birnstiel}, T. 2023, arXiv e-prints, arXiv:2312.13287,
  \dodoi{10.48550/arXiv.2312.13287}

\bibitem[{{Birnstiel} {et~al.}(2012){Birnstiel}, {Klahr}, \&
  {Ercolano}}]{Birnstiel_2012}
{Birnstiel}, T., {Klahr}, H., \& {Ercolano}, B. 2012, \aap, 539, A148,
  \dodoi{10.1051/0004-6361/201118136}

\bibitem[{{Birnstiel} {et~al.}(2018){Birnstiel}, {Dullemond}, {Zhu}, {Andrews},
  {Bai}, {Wilner}, {Carpenter}, {Huang}, {Isella}, {Benisty}, {P{\'e}rez}, \&
  {Zhang}}]{Birnstiel_2018}
{Birnstiel}, T., {Dullemond}, C.~P., {Zhu}, Z., {et~al.} 2018, \apjl, 869, L45,
  \dodoi{10.3847/2041-8213/aaf743}

\bibitem[{{Carrasco-Gonz{\'a}lez} {et~al.}(2019){Carrasco-Gonz{\'a}lez},
  {Sierra}, {Flock}, {Zhu}, {Henning}, {Chandler}, {Galv{\'a}n-Madrid},
  {Mac{\'\i}as}, {Anglada}, {Linz}, {Osorio}, {Rodr{\'\i}guez}, {Testi},
  {Torrelles}, {P{\'e}rez}, \& {Liu}}]{Carrasco-Gonzalez_2019}
{Carrasco-Gonz{\'a}lez}, C., {Sierra}, A., {Flock}, M., {et~al.} 2019, \apj,
  883, 71, \dodoi{10.3847/1538-4357/ab3d33}

\bibitem[{{Casassus} {et~al.}(2022){Casassus}, {C{\'a}rcamo}, {Hales}, {Weber},
  \& {Dent}}]{Casassus_2022}
{Casassus}, S., {C{\'a}rcamo}, M., {Hales}, A., {Weber}, P., \& {Dent}, B.
  2022, \apjl, 933, L4, \dodoi{10.3847/2041-8213/ac75e8}

\bibitem[{{Casassus} \& {P{\'e}rez}(2019)}]{Casassus_2019}
{Casassus}, S., \& {P{\'e}rez}, S. 2019, \apjl, 883, L41,
  \dodoi{10.3847/2041-8213/ab4425}

\bibitem[{{Currie} {et~al.}(2022){Currie}, {Lawson}, {Schneider}, {Lyra},
  {Wisniewski}, {Grady}, {Guyon}, {Tamura}, {Kotani}, {Kawahara}, {Brandt},
  {Uyama}, {Muto}, {Dong}, {Kudo}, {Hashimoto}, {Fukagawa}, {Wagner}, {Lozi},
  {Chilcote}, {Tobin}, {Groff}, {Ward-Duong}, {Januszewski}, {Norris},
  {Tuthill}, {van der Marel}, {Sitko}, {Deo}, {Vievard}, {Jovanovic},
  {Martinache}, \& {Skaf}}]{Currie_2022}
{Currie}, T., {Lawson}, K., {Schneider}, G., {et~al.} 2022, Nature Astronomy,
  6, 751, \dodoi{10.1038/s41550-022-01634-x}

\bibitem[{{Czekala} {et~al.}(2021){Czekala}, {Loomis}, {Teague}, {Booth},
  {Huang}, {Cataldi}, {Ilee}, {Law}, {Walsh}, {Bosman}, {Guzm{\'a}n}, {Le Gal},
  {{\"O}berg}, {Yamato}, {Aikawa}, {Andrews}, {Bae}, {Bergin}, {Bergner},
  {Cleeves}, {Kurtovic}, {M{\'e}nard}, {Nomura}, {P{\'e}rez}, {Qi}, {Schwarz},
  {Tsukagoshi}, {Waggoner}, {Wilner}, \& {Zhang}}]{Czekala_2021}
{Czekala}, I., {Loomis}, R.~A., {Teague}, R., {et~al.} 2021, \apjs, 257, 2,
  \dodoi{10.3847/1538-4365/ac1430}

\bibitem[{{Dipierro} \& {Laibe}(2017)}]{Dipierro_2017}
{Dipierro}, G., \& {Laibe}, G. 2017, \mnras, 469, 1932,
  \dodoi{10.1093/mnras/stx977}

\bibitem[{{Disk Dynamics Collaboration} {et~al.}(2020){Disk Dynamics
  Collaboration}, {Armitage}, {Bae}, {Benisty}, {Bergin}, {Casassus},
  {Czekala}, {Facchini}, {Fung}, {Hall}, {Ilee}, {Keppler}, {Kuznetsova}, {Le
  Gal}, {Loomis}, {Lyra}, {Manger}, {Perez}, {Pinte}, {Price}, {Rosotti},
  {Szulagyi}, {Schwarz}, {Simon}, {Teague}, \& {Zhang}}]{Armitage_2020}
{Disk Dynamics Collaboration}, {Armitage}, P.~J., {Bae}, J., {et~al.} 2020,
  arXiv e-prints, arXiv:2009.04345, \dodoi{10.48550/arXiv.2009.04345}

\bibitem[{{Doi} \& {Kataoka}(2021)}]{Kiyoaki_2021}
{Doi}, K., \& {Kataoka}, A. 2021, \apj, 912, 164,
  \dodoi{10.3847/1538-4357/abe5a6}

\bibitem[{{Dra{\.z}kowska} {et~al.}(2023){Dra{\.z}kowska}, {Bitsch},
  {Lambrechts}, {Mulders}, {Harsono}, {Vazan}, {Liu}, {Ormel}, {Kretke}, \&
  {Morbidelli}}]{Drkazkowska_2023}
{Dra{\.z}kowska}, J., {Bitsch}, B., {Lambrechts}, M., {et~al.} 2023, in
  Astronomical Society of the Pacific Conference Series, Vol. 534, Protostars
  and Planets VII, ed. S.~{Inutsuka}, Y.~{Aikawa}, T.~{Muto}, K.~{Tomida}, \&
  M.~{Tamura}, 717, \dodoi{10.48550/arXiv.2203.09759}

\bibitem[{{Dutrey} {et~al.}(2017){Dutrey}, {Guilloteau}, {Pi{\'e}tu},
  {Chapillon}, {Wakelam}, {Di Folco}, {Stoecklin}, {Denis-Alpizar}, {Gorti},
  {Teague}, {Henning}, {Semenov}, \& {Grosso}}]{Dutrey_2017}
{Dutrey}, A., {Guilloteau}, S., {Pi{\'e}tu}, V., {et~al.} 2017, \aap, 607,
  A130, \dodoi{10.1051/0004-6361/201730645}

\bibitem[{{Fang} {et~al.}(2023){Fang}, {Pascucci}, {Edwards}, {Gorti},
  {Hillenbrand}, \& {Carpenter}}]{Fang_2023}
{Fang}, M., {Pascucci}, I., {Edwards}, S., {et~al.} 2023, \apj, 945, 112,
  \dodoi{10.3847/1538-4357/acb2c9}

\bibitem[{{Flock} {et~al.}(2015){Flock}, {Ruge}, {Dzyurkevich}, {Henning},
  {Klahr}, \& {Wolf}}]{Flock_2015}
{Flock}, M., {Ruge}, J.~P., {Dzyurkevich}, N., {et~al.} 2015, \aap, 574, A68,
  \dodoi{10.1051/0004-6361/201424693}

\bibitem[{Foreman-Mackey(2016)}]{corner}
Foreman-Mackey, D. 2016, The Journal of Open Source Software, 1, 24,
  \dodoi{10.21105/joss.00024}

\bibitem[{{Foreman-Mackey} {et~al.}(2013){Foreman-Mackey}, {Hogg}, {Lang}, \&
  {Goodman}}]{Foreman_2013}
{Foreman-Mackey}, D., {Hogg}, D.~W., {Lang}, D., \& {Goodman}, J. 2013, \pasp,
  125, 306, \dodoi{10.1086/670067}

\bibitem[{{Gaia Collaboration} {et~al.}(2023){Gaia Collaboration}, {Vallenari},
  {Brown}, {Prusti}, {de Bruijne}, {Arenou}, {Babusiaux}, {Biermann},
  {Creevey}, {Ducourant}, {Evans}, {Eyer}, {Guerra}, {Hutton}, {Jordi},
  {Klioner}, {Lammers}, {Lindegren}, {Luri}, {Mignard}, {Panem}, {Pourbaix},
  {Randich}, {Sartoretti}, {Soubiran}, {Tanga}, {Walton}, {Bailer-Jones},
  {Bastian}, {Drimmel}, {Jansen}, {Katz}, {Lattanzi}, {van Leeuwen}, {Bakker},
  {Cacciari}, {Casta{\~n}eda}, {De Angeli}, {Fabricius}, {Fouesneau},
  {Fr{\'e}mat}, {Galluccio}, {Guerrier}, {Heiter}, {Masana}, {Messineo},
  {Mowlavi}, {Nicolas}, {Nienartowicz}, {Pailler}, {Panuzzo}, {Riclet}, {Roux},
  {Seabroke}, {Sordo}, {Th{\'e}venin}, {Gracia-Abril}, {Portell}, {Teyssier},
  {Altmann}, {Andrae}, {Audard}, {Bellas-Velidis}, {Benson}, {Berthier},
  {Blomme}, {Burgess}, {Busonero}, {Busso}, {C{\'a}novas}, {Carry}, {Cellino},
  {Cheek}, {Clementini}, {Damerdji}, {Davidson}, {de Teodoro}, {Nu{\~n}ez
  Campos}, {Delchambre}, {Dell'Oro}, {Esquej}, {Fern{\'a}ndez-Hern{\'a}ndez},
  {Fraile}, {Garabato}, {Garc{\'\i}a-Lario}, {Gosset}, {Haigron}, {Halbwachs},
  {Hambly}, {Harrison}, {Hern{\'a}ndez}, {Hestroffer}, {Hodgkin}, {Holl},
  {Jan{\ss}en}, {Jevardat de Fombelle}, {Jordan}, {Krone-Martins}, {Lanzafame},
  {L{\"o}ffler}, {Marchal}, {Marrese}, {Moitinho}, {Muinonen}, {Osborne},
  {Pancino}, {Pauwels}, {Recio-Blanco}, {Reyl{\'e}}, {Riello}, {Rimoldini},
  {Roegiers}, {Rybizki}, {Sarro}, {Siopis}, {Smith}, {Sozzetti}, {Utrilla},
  {van Leeuwen}, {Abbas}, {{\'A}brah{\'a}m}, {Abreu Aramburu}, {Aerts},
  {Aguado}, {Ajaj}, {Aldea-Montero}, {Altavilla}, {{\'A}lvarez}, {Alves},
  {Anders}, {Anderson}, {Anglada Varela}, {Antoja}, {Baines}, {Baker},
  {Balaguer-N{\'u}{\~n}ez}, {Balbinot}, {Balog}, {Barache}, {Barbato},
  {Barros}, {Barstow}, {Bartolom{\'e}}, {Bassilana}, {Bauchet}, {Becciani},
  {Bellazzini}, {Berihuete}, {Bernet}, {Bertone}, {Bianchi}, {Binnenfeld},
  {Blanco-Cuaresma}, {Blazere}, {Boch}, {Bombrun}, {Bossini}, {Bouquillon},
  {Bragaglia}, {Bramante}, {Breedt}, {Bressan}, {Brouillet}, {Brugaletta},
  {Bucciarelli}, {Burlacu}, {Butkevich}, {Buzzi}, {Caffau}, {Cancelliere},
  {Cantat-Gaudin}, {Carballo}, {Carlucci}, {Carnerero}, {Carrasco},
  {Casamiquela}, {Castellani}, {Castro-Ginard}, {Chaoul}, {Charlot}, {Chemin},
  {Chiaramida}, {Chiavassa}, {Chornay}, {Comoretto}, {Contursi}, {Cooper},
  {Cornez}, {Cowell}, {Crifo}, {Cropper}, {Crosta}, {Crowley}, {Dafonte},
  {Dapergolas}, {David}, {David}, {de Laverny}, {De Luise}, {De March}, {De
  Ridder}, {de Souza}, {de Torres}, {del Peloso}, {del Pozo}, {Delbo},
  {Delgado}, {Delisle}, {Demouchy}, {Dharmawardena}, {Di Matteo}, {Diakite},
  {Diener}, {Distefano}, {Dolding}, {Edvardsson}, {Enke}, {Fabre}, {Fabrizio},
  {Faigler}, {Fedorets}, {Fernique}, {Fienga}, {Figueras}, {Fournier},
  {Fouron}, {Fragkoudi}, {Gai}, {Garcia-Gutierrez}, {Garcia-Reinaldos},
  {Garc{\'\i}a-Torres}, {Garofalo}, {Gavel}, {Gavras}, {Gerlach}, {Geyer},
  {Giacobbe}, {Gilmore}, {Girona}, {Giuffrida}, {Gomel}, {Gomez},
  {Gonz{\'a}lez-N{\'u}{\~n}ez}, {Gonz{\'a}lez-Santamar{\'\i}a},
  {Gonz{\'a}lez-Vidal}, {Granvik}, {Guillout}, {Guiraud},
  {Guti{\'e}rrez-S{\'a}nchez}, {Guy}, {Hatzidimitriou}, {Hauser}, {Haywood},
  {Helmer}, {Helmi}, {Sarmiento}, {Hidalgo}, {Hilger}, {H{\l}adczuk}, {Hobbs},
  {Holland}, {Huckle}, {Jardine}, {Jasniewicz}, {Jean-Antoine Piccolo},
  {Jim{\'e}nez-Arranz}, {Jorissen}, {Juaristi Campillo}, {Julbe}, {Karbevska},
  {Kervella}, {Khanna}, {Kontizas}, {Kordopatis}, {Korn}, {K{\'o}sp{\'a}l},
  {Kostrzewa-Rutkowska}, {Kruszy{\'n}ska}, {Kun}, {Laizeau}, {Lambert},
  {Lanza}, {Lasne}, {Le Campion}, {Lebreton}, {Lebzelter}, {Leccia}, {Leclerc},
  {Lecoeur-Taibi}, {Liao}, {Licata}, {Lindstr{\o}m}, {Lister}, {Livanou},
  {Lobel}, {Lorca}, {Loup}, {Madrero Pardo}, {Magdaleno Romeo}, {Managau},
  {Mann}, {Manteiga}, {Marchant}, {Marconi}, {Marcos}, {Marcos Santos},
  {Mar{\'\i}n Pina}, {Marinoni}, {Marocco}, {Marshall}, {Martin Polo},
  {Mart{\'\i}n-Fleitas}, {Marton}, {Mary}, {Masip}, {Massari},
  {Mastrobuono-Battisti}, {Mazeh}, {McMillan}, {Messina}, {Michalik}, {Millar},
  {Mints}, {Molina}, {Molinaro}, {Moln{\'a}r}, {Monari}, {Mongui{\'o}},
  {Montegriffo}, {Montero}, {Mor}, {Mora}, {Morbidelli}, {Morel}, {Morris},
  {Muraveva}, {Murphy}, {Musella}, {Nagy}, {Noval}, {Oca{\~n}a}, {Ogden},
  {Ordenovic}, {Osinde}, {Pagani}, {Pagano}, {Palaversa}, {Palicio},
  {Pallas-Quintela}, {Panahi}, {Payne-Wardenaar}, {Pe{\~n}alosa Esteller},
  {Penttil{\"a}}, {Pichon}, {Piersimoni}, {Pineau}, {Plachy}, {Plum}, {Poggio},
  {Pr{\v{s}}a}, {Pulone}, {Racero}, {Ragaini}, {Rainer}, {Raiteri}, {Rambaux},
  {Ramos}, {Ramos-Lerate}, {Re Fiorentin}, {Regibo}, {Richards}, {Rios Diaz},
  {Ripepi}, {Riva}, {Rix}, {Rixon}, {Robichon}, {Robin}, {Robin}, {Roelens},
  {Rogues}, {Rohrbasser}, {Romero-G{\'o}mez}, {Rowell}, {Royer}, {Ruz Mieres},
  {Rybicki}, {Sadowski}, {S{\'a}ez N{\'u}{\~n}ez}, {Sagrist{\`a} Sell{\'e}s},
  {Sahlmann}, {Salguero}, {Samaras}, {Sanchez Gimenez}, {Sanna},
  {Santove{\~n}a}, {Sarasso}, {Schultheis}, {Sciacca}, {Segol}, {Segovia},
  {S{\'e}gransan}, {Semeux}, {Shahaf}, {Siddiqui}, {Siebert}, {Siltala},
  {Silvelo}, {Slezak}, {Slezak}, {Smart}, {Snaith}, {Solano}, {Solitro},
  {Souami}, {Souchay}, {Spagna}, {Spina}, {Spoto}, {Steele},
  {Steidelm{\"u}ller}, {Stephenson}, {S{\"u}veges}, {Surdej}, {Szabados},
  {Szegedi-Elek}, {Taris}, {Taylor}, {Teixeira}, {Tolomei}, {Tonello}, {Torra},
  {Torra}, {Torralba Elipe}, {Trabucchi}, {Tsounis}, {Turon}, {Ulla}, {Unger},
  {Vaillant}, {van Dillen}, {van Reeven}, {Vanel}, {Vecchiato}, {Viala},
  {Vicente}, {Voutsinas}, {Weiler}, {Wevers}, {Wyrzykowski}, {Yoldas}, {Yvard},
  {Zhao}, {Zorec}, {Zucker}, \& {Zwitter}}]{Gaia_2023}
{Gaia Collaboration}, {Vallenari}, A., {Brown}, A.~G.~A., {et~al.} 2023, \aap,
  674, A1, \dodoi{10.1051/0004-6361/202243940}

\bibitem[{{Galloway-Sprietsma} {et~al.}(2023){Galloway-Sprietsma}, {Bae},
  {Teague}, {Benisty}, {Facchini}, {Aikawa}, {Alarc{\'o}n}, {Andrews},
  {Bergin}, {Cataldi}, {Cleeves}, {Czekala}, {Guzm{\'a}n}, {Huang}, {Law}, {Le
  Gal}, {Liu}, {Long}, {M{\'e}nard}, {{\"O}berg}, {Walsh}, \&
  {Wilner}}]{Galloway_2023}
{Galloway-Sprietsma}, M., {Bae}, J., {Teague}, R., {et~al.} 2023, \apj, 950,
  147, \dodoi{10.3847/1538-4357/accae4}

\bibitem[{{G{\'a}rate} {et~al.}(2021){G{\'a}rate}, {Delage}, {Stadler},
  {Pinilla}, {Birnstiel}, {Stammler}, {Picogna}, {Ercolano}, {Franz}, \&
  {Lenz}}]{Garate_2021}
{G{\'a}rate}, M., {Delage}, T.~N., {Stadler}, J., {et~al.} 2021, \aap, 655,
  A18, \dodoi{10.1051/0004-6361/202141444}

\bibitem[{{G{\'a}rate} {et~al.}(2023){G{\'a}rate}, {Birnstiel}, {Pinilla},
  {Andrews}, {Franz}, {Stammler}, {Picogna}, {Ercolano}, {Miotello}, \&
  {Kurtovic}}]{Garate_2023}
{G{\'a}rate}, M., {Birnstiel}, T., {Pinilla}, P., {et~al.} 2023, \aap, 679,
  A15, \dodoi{10.1051/0004-6361/202244436}

\bibitem[{{Goldreich} \& {Ward}(1973)}]{Goldreich_1973}
{Goldreich}, P., \& {Ward}, W.~R. 1973, \apj, 183, 1051, \dodoi{10.1086/152291}

\bibitem[{{Haffert} {et~al.}(2019){Haffert}, {Bohn}, {de Boer}, {Snellen},
  {Brinchmann}, {Girard}, {Keller}, \& {Bacon}}]{Haffert_2019}
{Haffert}, S.~Y., {Bohn}, A.~J., {de Boer}, J., {et~al.} 2019, Nature
  Astronomy, 3, 749, \dodoi{10.1038/s41550-019-0780-5}

\bibitem[{Harris {et~al.}(2020)Harris, Millman, van~der Walt, Gommers,
  Virtanen, Cournapeau, Wieser, Taylor, Berg, Smith, Kern, Picus, Hoyer, van
  Kerkwijk, Brett, Haldane, del R{\'{i}}o, Wiebe, Peterson,
  G{\'{e}}rard-Marchant, Sheppard, Reddy, Weckesser, Abbasi, Gohlke, \&
  Oliphant}]{Numpy_2020}
Harris, C.~R., Millman, K.~J., van~der Walt, S.~J., {et~al.} 2020, Nature, 585,
  357, \dodoi{10.1038/s41586-020-2649-2}

\bibitem[{{Huang} {et~al.}(2018){Huang}, {Andrews}, {Dullemond}, {Isella},
  {P{\'e}rez}, {Guzm{\'a}n}, {{\"O}berg}, {Zhu}, {Zhang}, {Bai}, {Benisty},
  {Birnstiel}, {Carpenter}, {Hughes}, {Ricci}, {Weaver}, \&
  {Wilner}}]{Huang_2018}
{Huang}, J., {Andrews}, S.~M., {Dullemond}, C.~P., {et~al.} 2018, \apjl, 869,
  L42, \dodoi{10.3847/2041-8213/aaf740}

\bibitem[{Hunter(2007)}]{Matplotlib_2007}
Hunter, J.~D. 2007, Computing in Science \& Engineering, 9, 90,
  \dodoi{10.1109/MCSE.2007.55}

\bibitem[{{Isella} {et~al.}(2019{\natexlab{a}}){Isella}, {Benisty}, {Teague},
  {Bae}, {Keppler}, {Facchini}, \& {P{\'e}rez}}]{Isella_2019b}
{Isella}, A., {Benisty}, M., {Teague}, R., {et~al.} 2019{\natexlab{a}}, \apjl,
  879, L25, \dodoi{10.3847/2041-8213/ab2a12}

\bibitem[{{Isella} {et~al.}(2019{\natexlab{b}}){Isella}, {Andrews},
  {Dullemond}, {P{\'e}rez}, {Huang}, {Birnstiel}, {Guzman}, {Kurtovic},
  {Zhang}, {Zhu}, {Bay}, {Benisty}, {Carpenter}, {Hughes}, {Oberg}, {Ricci},
  {Weaver}, \& {Wilner}}]{Isella_2019}
{Isella}, A., {Andrews}, S.~M., {Dullemond}, C.~P., {et~al.}
  2019{\natexlab{b}}, in 50th Annual Lunar and Planetary Science Conference,
  Lunar and Planetary Science Conference, 2821

\bibitem[{{Izquierdo} {et~al.}(2021){Izquierdo}, {Testi}, {Facchini},
  {Rosotti}, \& {van Dishoeck}}]{Izquierdo_2021}
{Izquierdo}, A.~F., {Testi}, L., {Facchini}, S., {Rosotti}, G.~P., \& {van
  Dishoeck}, E.~F. 2021, \aap, 650, A179, \dodoi{10.1051/0004-6361/202140779}

\bibitem[{{Jennings} {et~al.}(2022){Jennings}, {Booth}, {Tazzari}, {Clarke}, \&
  {Rosotti}}]{Jennings_2021}
{Jennings}, J., {Booth}, R.~A., {Tazzari}, M., {Clarke}, C.~J., \& {Rosotti},
  G.~P. 2022, \mnras, 509, 2780, \dodoi{10.1093/mnras/stab3185}

\bibitem[{{Jennings} {et~al.}(2020){Jennings}, {Booth}, {Tazzari}, {Rosotti},
  \& {Clarke}}]{Jennings_2020}
{Jennings}, J., {Booth}, R.~A., {Tazzari}, M., {Rosotti}, G.~P., \& {Clarke},
  C.~J. 2020, \mnras, 495, 3209, \dodoi{10.1093/mnras/staa1365}

\bibitem[{{Jorsater} \& {van Moorsel}(1995)}]{Jorsater_1995}
{Jorsater}, S., \& {van Moorsel}, G.~A. 1995, \aj, 110, 2037,
  \dodoi{10.1086/117668}

\bibitem[{{Keppler} {et~al.}(2018){Keppler}, {Benisty}, {M{\"u}ller},
  {Henning}, {van Boekel}, {Cantalloube}, {Ginski}, {van Holstein}, {Maire},
  {Pohl}, {Samland}, {Avenhaus}, {Baudino}, {Boccaletti}, {de Boer},
  {Bonnefoy}, {Chauvin}, {Desidera}, {Langlois}, {Lazzoni}, {Marleau},
  {Mordasini}, {Pawellek}, {Stolker}, {Vigan}, {Zurlo}, {Birnstiel},
  {Brandner}, {Feldt}, {Flock}, {Girard}, {Gratton}, {Hagelberg}, {Isella},
  {Janson}, {Juhasz}, {Kemmer}, {Kral}, {Lagrange}, {Launhardt}, {Matter},
  {M{\'e}nard}, {Milli}, {Molli{\`e}re}, {Olofsson}, {P{\'e}rez}, {Pinilla},
  {Pinte}, {Quanz}, {Schmidt}, {Udry}, {Wahhaj}, {Williams}, {Buenzli},
  {Cudel}, {Dominik}, {Galicher}, {Kasper}, {Lannier}, {Mesa}, {Mouillet},
  {Peretti}, {Perrot}, {Salter}, {Sissa}, {Wildi}, {Abe}, {Antichi},
  {Augereau}, {Baruffolo}, {Baudoz}, {Bazzon}, {Beuzit}, {Blanchard}, {Brems},
  {Buey}, {De Caprio}, {Carbillet}, {Carle}, {Cascone}, {Cheetham}, {Claudi},
  {Costille}, {Delboulb{\'e}}, {Dohlen}, {Fantinel}, {Feautrier}, {Fusco},
  {Giro}, {Gluck}, {Gry}, {Hubin}, {Hugot}, {Jaquet}, {Le Mignant}, {Llored},
  {Madec}, {Magnard}, {Martinez}, {Maurel}, {Meyer}, {M{\"o}ller-Nilsson},
  {Moulin}, {Mugnier}, {Orign{\'e}}, {Pavlov}, {Perret}, {Petit}, {Pragt},
  {Puget}, {Rabou}, {Ramos}, {Rigal}, {Rochat}, {Roelfsema}, {Rousset}, {Roux},
  {Salasnich}, {Sauvage}, {Sevin}, {Soenke}, {Stadler}, {Suarez}, {Turatto}, \&
  {Weber}}]{Keppler_2018}
{Keppler}, M., {Benisty}, M., {M{\"u}ller}, A., {et~al.} 2018, \aap, 617, A44,
  \dodoi{10.1051/0004-6361/201832957}

\bibitem[{{Kurtovic} {et~al.}(2022){Kurtovic}, {Pinilla}, {Penzlin}, {Benisty},
  {P{\'e}rez}, {Ginski}, {Isella}, {Kley}, {Menard}, {P{\'e}rez}, \&
  {Bayo}}]{Kurtovic_2022}
{Kurtovic}, N.~T., {Pinilla}, P., {Penzlin}, A. B.~T., {et~al.} 2022, \aap,
  664, A151, \dodoi{10.1051/0004-6361/202243505}

\bibitem[{{Lada}(1987)}]{Lada_1987}
{Lada}, C.~J. 1987, in Star Forming Regions, ed. M.~{Peimbert} \& J.~{Jugaku},
  Vol. 115, 1

\bibitem[{{Law} {et~al.}(2021{\natexlab{a}}){Law}, {Loomis}, {Teague},
  {{\"O}berg}, {Czekala}, {Andrews}, {Huang}, {Aikawa}, {Alarc{\'o}n}, {Bae},
  {Bergin}, {Bergner}, {Boehler}, {Booth}, {Bosman}, {Calahan}, {Cataldi},
  {Cleeves}, {Furuya}, {Guzm{\'a}n}, {Ilee}, {Le Gal}, {Liu}, {Long},
  {M{\'e}nard}, {Nomura}, {Qi}, {Schwarz}, {Sierra}, {Tsukagoshi}, {Yamato},
  {van't Hoff}, {Walsh}, {Wilner}, \& {Zhang}}]{Law_2020_rad}
{Law}, C.~J., {Loomis}, R.~A., {Teague}, R., {et~al.} 2021{\natexlab{a}},
  \apjs, 257, 3, \dodoi{10.3847/1538-4365/ac1434}

\bibitem[{{Law} {et~al.}(2021{\natexlab{b}}){Law}, {Teague}, {Loomis}, {Bae},
  {{\"O}berg}, {Czekala}, {Andrews}, {Aikawa}, {Alarc{\'o}n}, {Bergin},
  {Bergner}, {Booth}, {Bosman}, {Calahan}, {Cataldi}, {Cleeves}, {Furuya},
  {Guzm{\'a}n}, {Huang}, {Ilee}, {Le Gal}, {Liu}, {Long}, {M{\'e}nard},
  {Nomura}, {P{\'e}rez}, {Qi}, {Schwarz}, {Soto}, {Tsukagoshi}, {Yamato},
  {van't Hoff}, {Walsh}, {Wilner}, \& {Zhang}}]{Law_2021}
{Law}, C.~J., {Teague}, R., {Loomis}, R.~A., {et~al.} 2021{\natexlab{b}},
  \apjs, 257, 4, \dodoi{10.3847/1538-4365/ac1439}

\bibitem[{{Long} {et~al.}(2018){Long}, {Pinilla}, {Herczeg}, {Harsono},
  {Dipierro}, {Pascucci}, {Hendler}, {Tazzari}, {Ragusa}, {Salyk}, {Edwards},
  {Lodato}, {van de Plas}, {Johnstone}, {Liu}, {Boehler}, {Cabrit}, {Manara},
  {Menard}, {Mulders}, {Nisini}, {Fischer}, {Rigliaco}, {Banzatti}, {Avenhaus},
  \& {Gully-Santiago}}]{Long_2018}
{Long}, F., {Pinilla}, P., {Herczeg}, G.~J., {et~al.} 2018, \apj, 869, 17,
  \dodoi{10.3847/1538-4357/aae8e1}

\bibitem[{{Long} {et~al.}(2022){Long}, {Andrews}, {Rosotti}, {Harsono},
  {Pinilla}, {Wilner}, {{\"O}berg}, {Teague}, {Trapman}, \&
  {Tabone}}]{Long_2022b}
{Long}, F., {Andrews}, S.~M., {Rosotti}, G., {et~al.} 2022, \apj, 931, 6,
  \dodoi{10.3847/1538-4357/ac634e}

\bibitem[{{McMullin} {et~al.}(2007){McMullin}, {Waters}, {Schiebel}, {Young},
  \& {Golap}}]{McMullin_2007}
{McMullin}, J.~P., {Waters}, B., {Schiebel}, D., {Young}, W., \& {Golap}, K.
  2007, in Astronomical Society of the Pacific Conference Series, Vol. 376,
  Astronomical Data Analysis Software and Systems XVI, ed. R.~A. {Shaw},
  F.~{Hill}, \& D.~J. {Bell}, 127

\bibitem[{{{\"O}berg} {et~al.}(2021){{\"O}berg}, {Guzm{\'a}n}, {Walsh},
  {Aikawa}, {Bergin}, {Law}, {Loomis}, {Alarc{\'o}n}, {Andrews}, {Bae},
  {Bergner}, {Boehler}, {Booth}, {Bosman}, {Calahan}, {Cataldi}, {Cleeves},
  {Czekala}, {Furuya}, {Huang}, {Ilee}, {Kurtovic}, {Le Gal}, {Liu}, {Long},
  {M{\'e}nard}, {Nomura}, {P{\'e}rez}, {Qi}, {Schwarz}, {Sierra}, {Teague},
  {Tsukagoshi}, {Yamato}, {van't Hoff}, {Waggoner}, {Wilner}, \&
  {Zhang}}]{Oberg_2021}
{{\"O}berg}, K.~I., {Guzm{\'a}n}, V.~V., {Walsh}, C., {et~al.} 2021, \apjs,
  257, 1, \dodoi{10.3847/1538-4365/ac1432}

\bibitem[{{Ohashi} {et~al.}(2023){Ohashi}, {Tobin}, {J{\o}rgensen}, {Takakuwa},
  {Sheehan}, {Aikawa}, {Li}, {Looney}, {Williams}, {Aso}, {Sharma}, {Sai},
  {Yamato}, {Lee}, {Tomida}, {Yen}, {Encalada}, {Flores}, {Gavino}, {Kido},
  {Han}, {Lin}, {Narayanan}, {Phuong}, {Santamar{\'\i}a-Miranda}, {Thieme},
  {van't Hoff}, {de Gregorio-Monsalvo}, {Koch}, {Kwon}, {Lai}, {Lee},
  {Plunkett}, {Saigo}, {Hirano}, {Lam}, \& {Mori}}]{Ohashi_2023}
{Ohashi}, N., {Tobin}, J.~J., {J{\o}rgensen}, J.~K., {et~al.} 2023, \apj, 951,
  8, \dodoi{10.3847/1538-4357/acd384}

\bibitem[{{Okuzumi} {et~al.}(2016){Okuzumi}, {Momose}, {Sirono}, {Kobayashi},
  \& {Tanaka}}]{Okuzumi_2016}
{Okuzumi}, S., {Momose}, M., {Sirono}, S.-i., {Kobayashi}, H., \& {Tanaka}, H.
  2016, \apj, 821, 82, \dodoi{10.3847/0004-637X/821/2/82}

\bibitem[{{Paneque-Carre{\~n}o} {et~al.}(2023){Paneque-Carre{\~n}o},
  {Miotello}, {van Dishoeck}, {Tabone}, {Izquierdo}, \&
  {Facchini}}]{Paneque_2023}
{Paneque-Carre{\~n}o}, T., {Miotello}, A., {van Dishoeck}, E.~F., {et~al.}
  2023, \aap, 669, A126, \dodoi{10.1051/0004-6361/202244428}

\bibitem[{{P{\'e}rez} {et~al.}(2016){P{\'e}rez}, {Carpenter}, {Andrews},
  {Ricci}, {Isella}, {Linz}, {Sargent}, {Wilner}, {Henning}, {Deller},
  {Chandler}, {Dullemond}, {Lazio}, {Menten}, {Corder}, {Storm}, {Testi},
  {Tazzari}, {Kwon}, {Calvet}, {Greaves}, {Harris}, \& {Mundy}}]{Perez_2016}
{P{\'e}rez}, L.~M., {Carpenter}, J.~M., {Andrews}, S.~M., {et~al.} 2016,
  Science, 353, 1519, \dodoi{10.1126/science.aaf8296}

\bibitem[{{P{\'e}rez} {et~al.}(2018){P{\'e}rez}, {Casassus}, \&
  {Ben{\'\i}tez-Llambay}}]{Perez_2018}
{P{\'e}rez}, S., {Casassus}, S., \& {Ben{\'\i}tez-Llambay}, P. 2018, \mnras,
  480, L12, \dodoi{10.1093/mnrasl/sly109}

\bibitem[{{Perez} {et~al.}(2015){Perez}, {Dunhill}, {Casassus}, {Roman},
  {Szul{\'a}gyi}, {Flores}, {Marino}, \& {Montesinos}}]{Perez_2015}
{Perez}, S., {Dunhill}, A., {Casassus}, S., {et~al.} 2015, \apjl, 811, L5,
  \dodoi{10.1088/2041-8205/811/1/L5}

\bibitem[{{Pinilla} {et~al.}(2012{\natexlab{a}}){Pinilla}, {Benisty}, \&
  {Birnstiel}}]{Pinilla_2012}
{Pinilla}, P., {Benisty}, M., \& {Birnstiel}, T. 2012{\natexlab{a}}, \aap, 545,
  A81, \dodoi{10.1051/0004-6361/201219315}

\bibitem[{{Pinilla} {et~al.}(2012{\natexlab{b}}){Pinilla}, {Birnstiel},
  {Ricci}, {Dullemond}, {Uribe}, {Testi}, \& {Natta}}]{Pinilla_2012b}
{Pinilla}, P., {Birnstiel}, T., {Ricci}, L., {et~al.} 2012{\natexlab{b}}, \aap,
  538, A114, \dodoi{10.1051/0004-6361/201118204}

\bibitem[{{Pinilla} {et~al.}(2014){Pinilla}, {Benisty}, {Birnstiel}, {Ricci},
  {Isella}, {Natta}, {Dullemond}, {Quiroga-Nu{\~n}ez}, {Henning}, \&
  {Testi}}]{Pinilla_2014}
{Pinilla}, P., {Benisty}, M., {Birnstiel}, T., {et~al.} 2014, \aap, 564, A51,
  \dodoi{10.1051/0004-6361/201323322}

\bibitem[{{Pinte} {et~al.}(2018{\natexlab{a}}){Pinte}, {Price}, {M{\'e}nard},
  {Duch{\^e}ne}, {Dent}, {Hill}, {de Gregorio-Monsalvo}, {Hales}, \&
  {Mentiplay}}]{Pinte_2018}
{Pinte}, C., {Price}, D.~J., {M{\'e}nard}, F., {et~al.} 2018{\natexlab{a}},
  \apjl, 860, L13, \dodoi{10.3847/2041-8213/aac6dc}

\bibitem[{{Pinte} {et~al.}(2018{\natexlab{b}}){Pinte}, {M{\'e}nard},
  {Duch{\^e}ne}, {Hill}, {Dent}, {Woitke}, {Maret}, {van der Plas}, {Hales},
  {Kamp}, {Thi}, {de Gregorio-Monsalvo}, {Rab}, {Quanz}, {Avenhaus}, {Carmona},
  \& {Casassus}}]{Pinte_disksurf_2018}
{Pinte}, C., {M{\'e}nard}, F., {Duch{\^e}ne}, G., {et~al.} 2018{\natexlab{b}},
  \aap, 609, A47, \dodoi{10.1051/0004-6361/201731377}

\bibitem[{{Pinte} {et~al.}(2019){Pinte}, {van der Plas}, {M{\'e}nard}, {Price},
  {Christiaens}, {Hill}, {Mentiplay}, {Ginski}, {Choquet}, {Boehler},
  {Duch{\^e}ne}, {Perez}, \& {Casassus}}]{Pinte_2019}
{Pinte}, C., {van der Plas}, G., {M{\'e}nard}, F., {et~al.} 2019, Nature
  Astronomy, 3, 1109, \dodoi{10.1038/s41550-019-0852-6}

\bibitem[{{Pinte} {et~al.}(2020){Pinte}, {Price}, {M{\'e}nard}, {Duch{\^e}ne},
  {Christiaens}, {Andrews}, {Huang}, {Hill}, {van der Plas}, {Perez}, {Isella},
  {Boehler}, {Dent}, {Mentiplay}, \& {Loomis}}]{Pinte_2020}
{Pinte}, C., {Price}, D.~J., {M{\'e}nard}, F., {et~al.} 2020, \apjl, 890, L9,
  \dodoi{10.3847/2041-8213/ab6dda}

\bibitem[{{Pinte} {et~al.}(2023){Pinte}, {Hammond}, {Price}, {Christiaens},
  {Andrews}, {Chauvin}, {P{\'e}rez}, {Jorquera}, {Garg}, {Norfolk}, {Calcino},
  \& {Bonnefoy}}]{Pinte_2023}
{Pinte}, C., {Hammond}, I., {Price}, D.~J., {et~al.} 2023, \mnras, 526, L41,
  \dodoi{10.1093/mnrasl/slad010}

\bibitem[{{Price} {et~al.}(2018){Price}, {Cuello}, {Pinte}, {Mentiplay},
  {Casassus}, {Christiaens}, {Kennedy}, {Cuadra}, {Sebastian Perez}, {Marino},
  {Armitage}, {Zurlo}, {Juhasz}, {Ragusa}, {Laibe}, \& {Lodato}}]{Price_2018}
{Price}, D.~J., {Cuello}, N., {Pinte}, C., {et~al.} 2018, \mnras, 477, 1270,
  \dodoi{10.1093/mnras/sty647}

\bibitem[{{Rabago} \& {Zhu}(2021)}]{Rabago_2021}
{Rabago}, I., \& {Zhu}, Z. 2021, \mnras, 502, 5325,
  \dodoi{10.1093/mnras/stab447}

\bibitem[{{Reg{\'a}ly} {et~al.}(2012){Reg{\'a}ly}, {Juh{\'a}sz}, {S{\'a}ndor},
  \& {Dullemond}}]{Regaly_2012}
{Reg{\'a}ly}, Z., {Juh{\'a}sz}, A., {S{\'a}ndor}, Z., \& {Dullemond}, C.~P.
  2012, \mnras, 419, 1701, \dodoi{10.1111/j.1365-2966.2011.19834.x}

\bibitem[{{Rosotti} {et~al.}(2020){Rosotti}, {Teague}, {Dullemond}, {Booth}, \&
  {Clarke}}]{Rosotti_2020}
{Rosotti}, G.~P., {Teague}, R., {Dullemond}, C., {Booth}, R.~A., \& {Clarke},
  C.~J. 2020, \mnras, 495, 173, \dodoi{10.1093/mnras/staa1170}

\bibitem[{{Segura-Cox} {et~al.}(2020){Segura-Cox}, {Schmiedeke}, {Pineda},
  {Stephens}, {Fern{\'a}ndez-L{\'o}pez}, {Looney}, {Caselli}, {Li}, {Mundy},
  {Kwon}, \& {Harris}}]{Segura-Cox_2020}
{Segura-Cox}, D.~M., {Schmiedeke}, A., {Pineda}, J.~E., {et~al.} 2020, \nat,
  586, 228, \dodoi{10.1038/s41586-020-2779-6}

\bibitem[{{Sierra} \& {Lizano}(2020)}]{Sierra_2020}
{Sierra}, A., \& {Lizano}, S. 2020, \apj, 892, 136,
  \dodoi{10.3847/1538-4357/ab7d32}

\bibitem[{{Sierra} {et~al.}(2019){Sierra}, {Lizano}, {Mac{\'\i}as},
  {Carrasco-Gonz{\'a}lez}, {Osorio}, \& {Flock}}]{Sierra_2019}
{Sierra}, A., {Lizano}, S., {Mac{\'\i}as}, E., {et~al.} 2019, \apj, 876, 7,
  \dodoi{10.3847/1538-4357/ab1265}

\bibitem[{{Sierra} {et~al.}(2021){Sierra}, {P{\'e}rez}, {Zhang}, {Law},
  {Guzm{\'a}n}, {Qi}, {Bosman}, {{\"O}berg}, {Andrews}, {Long}, {Teague},
  {Booth}, {Walsh}, {Wilner}, {M{\'e}nard}, {Cataldi}, {Czekala}, {Bae},
  {Huang}, {Bergner}, {Ilee}, {Benisty}, {Le Gal}, {Loomis}, {Tsukagoshi},
  {Liu}, {Yamato}, \& {Aikawa}}]{Sierra_2021}
{Sierra}, A., {P{\'e}rez}, L.~M., {Zhang}, K., {et~al.} 2021, \apjs, 257, 14,
  \dodoi{10.3847/1538-4365/ac1431}

\bibitem[{{Takeuchi} \& {Lin}(2002)}]{Takeuchi_2002}
{Takeuchi}, T., \& {Lin}, D.~N.~C. 2002, \apj, 581, 1344,
  \dodoi{10.1086/344437}

\bibitem[{Tazzari(2017)}]{uvplot_tazzari}
Tazzari, M. 2017, mtazzari/uvplot,  Zenodo, \dodoi{10.5281/zenodo.1003113}

\bibitem[{{Tazzari} {et~al.}(2018){Tazzari}, {Beaujean}, \&
  {Testi}}]{Tazzari_2018}
{Tazzari}, M., {Beaujean}, F., \& {Testi}, L. 2018, \mnras, 476, 4527,
  \dodoi{10.1093/mnras/sty409}

\bibitem[{Teague(2019)}]{eddy}
Teague, R. 2019, The Journal of Open Source Software, 4, 1220,
  \dodoi{10.21105/joss.01220}

\bibitem[{{Teague} {et~al.}(2018){Teague}, {Bae}, {Birnstiel}, \&
  {Bergin}}]{Teague_2018}
{Teague}, R., {Bae}, J., {Birnstiel}, T., \& {Bergin}, E.~A. 2018, \apj, 868,
  113, \dodoi{10.3847/1538-4357/aae836}

\bibitem[{{Teague} {et~al.}(2019){Teague}, {Bae}, {Huang}, \&
  {Bergin}}]{Teague_2019}
{Teague}, R., {Bae}, J., {Huang}, J., \& {Bergin}, E.~A. 2019, \apjl, 884, L56,
  \dodoi{10.3847/2041-8213/ab4a83}

\bibitem[{{Teague} \& {Foreman-Mackey}(2018)}]{Teague_2018b}
{Teague}, R., \& {Foreman-Mackey}, D. 2018, Research Notes of the American
  Astronomical Society, 2, 173, \dodoi{10.3847/2515-5172/aae265}

\bibitem[{Teague {et~al.}(2021)Teague, Law, Huang, \& Meng}]{disksurf}
Teague, R., Law, C.~J., Huang, J., \& Meng, F. 2021, Journal of Open Source
  Software, 6, 3827, \dodoi{10.21105/joss.03827}

\bibitem[{{Teague} {et~al.}(2021){Teague}, {Bae}, {Aikawa}, {Andrews},
  {Bergin}, {Bergner}, {Boehler}, {Booth}, {Bosman}, {Cataldi}, {Czekala},
  {Guzm{\'a}n}, {Huang}, {Ilee}, {Law}, {Le Gal}, {Long}, {Loomis},
  {M{\'e}nard}, {{\"O}berg}, {P{\'e}rez}, {Schwarz}, {Sierra}, {Walsh},
  {Wilner}, {Yamato}, \& {Zhang}}]{Teague_2021}
{Teague}, R., {Bae}, J., {Aikawa}, Y., {et~al.} 2021, \apjs, 257, 18,
  \dodoi{10.3847/1538-4365/ac1438}

\bibitem[{{Trapman} {et~al.}(2019){Trapman}, {Facchini}, {Hogerheijde}, {van
  Dishoeck}, \& {Bruderer}}]{Trapman_2019}
{Trapman}, L., {Facchini}, S., {Hogerheijde}, M.~R., {van Dishoeck}, E.~F., \&
  {Bruderer}, S. 2019, \aap, 629, A79, \dodoi{10.1051/0004-6361/201834723}

\bibitem[{{Weidenschilling}(1977)}]{Weidenschilling_1977}
{Weidenschilling}, S.~J. 1977, \mnras, 180, 57, \dodoi{10.1093/mnras/180.2.57}

\bibitem[{{Williams} \& {Cieza}(2011)}]{Williams_2011}
{Williams}, J.~P., \& {Cieza}, L.~A. 2011, \araa, 49, 67,
  \dodoi{10.1146/annurev-astro-081710-102548}

\bibitem[{{Yu} {et~al.}(2021){Yu}, {Teague}, {Bae}, \& {{\"O}berg}}]{Yu_2021}
{Yu}, H., {Teague}, R., {Bae}, J., \& {{\"O}berg}, K. 2021, \apjl, 920, L33,
  \dodoi{10.3847/2041-8213/ac283e}

\bibitem[{{Zhang} {et~al.}(2015){Zhang}, {Blake}, \& {Bergin}}]{Zhang_2015}
{Zhang}, K., {Blake}, G.~A., \& {Bergin}, E.~A. 2015, \apjl, 806, L7,
  \dodoi{10.1088/2041-8205/806/1/L7}

\bibitem[{{Zhao} {et~al.}(2024){Zhao}, {Yu}, \& {Li}}]{Zhao_2024}
{Zhao}, M., {Yu}, H., \& {Li}, Z. 2024, Research in Astronomy and Astrophysics,
  24, 065010, \dodoi{10.1088/1674-4527/ad3ec7}

\end{thebibliography}
\bibliographystyle{aasjournal}

\end{document}